\documentclass[preprint2]{aastex6}
\usepackage{natbib}
\usepackage{graphicx}
\usepackage{amsmath}
\usepackage{amssymb}
\usepackage{longtable}

\newcommand{\Balmerint}{({\rm H}\alpha/{\rm H}\beta)_{\rm int}}
\newcommand{\Balmerobs}{({\rm H}\alpha/{\rm H}\beta)_{\rm obs}}

\newcommand{\ebv}{E_{\rm B-V}}

\newcommand{\kev}{\,\rm keV}
\newcommand{\cloudy}{{\sc cloudy}}

\newcommand{\NNaI}{N_{\rm NaI}}
\newcommand{\fNaI}{f_{\rm NaI}}
\newcommand{\NNa}{N_{\rm Na}}
\newcommand{\NH}{N_{\rm H}}
\newcommand{\cm}{\,{\rm cm}}

\newcommand{\Ha}{\text{H$\alpha$}}
\newcommand{\Hb}{\text{H$\beta$}}
\newcommand{\pc}{\,{\rm pc}}
\newcommand{\kpc}{\,{\rm kpc}}
\newcommand{\msun}{\,{\rm M_{\odot}}}
\newcommand{\mic}{\,\mbox{$\mu$m}}
\newcommand{\rdust}{r_{\rm d}}

\newcommand{\Ledd}{L_{\rm Edd}}
\newcommand{\mbh}{M_{\rm BH}}
\newcommand{\ergs}{\,\rm erg\ s^{-1}}

\newcommand{\K}{\,{\rm K}}
\newcommand{\lbol}{L_{\rm bol}}
\newcommand{\aopt}{{\alpha_{\rm opt}}}
\newcommand{\kms}{\,\rm km\ s^{-1}}
\newcommand{\oiii}{\text{[O~{\sc iii}]}}

\newcommand{\Sdust}{\Sigma_{\rm d}}
\newcommand{\Sgas}{\Sigma_{\rm g}}
\newcommand{\mP}{m_{\rm p}}

\newcommand{\fhost}{f_{\rm host}}
\renewcommand{\mag}{\,{\rm mag}}
\let\oldAA\AA
\renewcommand{\AA}{\text{\normalfont\oldAA}}

\begin{document}

\title {Evidence that most type 1 AGN are reddened by dust in the host ISM}
\shorttitle{evidence that type 1 AGN are reddened}
\shortauthors{Baron et al.}

\author{Dalya Baron\altaffilmark{1}, 
Jonathan Stern\altaffilmark{2,3}, 
Dovi Poznanski\altaffilmark{1}, and
Hagai Netzer\altaffilmark{1}}

\email{dalyabaron@mail.tau.ac.il}

\altaffiltext{1}{School of Physics and Astronomy, Tel-Aviv University, Tel Aviv 69978, Israel.}
\altaffiltext{2}{Max Planck Institut f$\ddot{u}$r Astronomie, K$\ddot{o}$nigstuhl 17, D-69117, Heidelberg, Germany.}
\altaffiltext{3}{Alexander von Humboldt fellow.}

\begin{abstract}
The typical optical-UV continuum slopes observed in many type 1 active galactic nuclei (AGN) are redder than expected from thin accretion disk models. A possible resolution to this conundrum is that many AGN are reddened by dust along the line of sight.
To explore this possibility, we stack $5000$ SDSS AGN with luminosity $L\approx10^{45}\ergs$ and redshift $z\sim0.4$ in bins of optical continuum slope $\aopt$ and width of the broad H$\beta$ emission line. We measure the equivalent width (EW) of the NaID absorption feature in each stacked spectrum. 
We find a linear relation between $\aopt$ and EW(NaID), such that EW(NaID) increases as $\aopt$ becomes redder. 
In the bin with the smallest H$\beta$ width, objects with the bluest slopes that are similar to accretion disk predictions are found to have ${\rm EW(NaID)} = 0$, supporting the line-of-sight dust hypothesis.
This conclusion is also supported by the dependence of the $\Ha/\Hb$ line ratio on $\aopt$.
The implied relationship between continuum slope and dust reddening is given by $\ebv \approx 0.2\cdot(-0.1 - \aopt)$, and the implied reddening of a typical type 1 AGN with $\aopt=-0.5$ is $\ebv\approx0.08\mag$.
Photoionization calculations show that the line-of-sight dusty gas responsible for reddening is too ionized to produce the observed sodium features. Therefore, we argue that the sodium absorption arises in regions of the host ISM which are shielded from the AGN radiation, along lines-of-sight to the stars, and the correlation with $\aopt$ arises since ISM columns along shielded and non-shielded sightlines are correlated. This scenario is supported by the similarity of the relation between $\ebv$ and NaI column implied by our results with the relation in the Milky-Way found by previous studies.

\end{abstract}


\vspace{1cm}
\section{Introduction}\label{s:intro}

The unified model of active galactic nuclei (AGN) consists of a central engine, believed to be an accretion disk around a black hole with mass $\mbh=10^6-10^{10}\msun$, and dust and gas that surround it \citep{antonucci93, urry95}. While many details have been added over the years (for a recent review see \citealt{netzer15}), this picture still gives an adequate description of the central part of galaxies containing active black holes (BHs) and the geometry in their centers. According to this picture, the observed spectrum of an AGN depends on the viewing angle relative to the distribution of the dusty nuclear gas. 
Broadly speaking, AGN are classified into type 1 (unobscured) and type 2 (obscured) AGN. 
In type 1 AGN, the line of sight provides a direct view of the dust-free gas close to the center that contain the central disk and fast moving high density clouds, 
while for type 2 AGN the emission from the accretion disk and inner gas clouds is completely obscured by dust. 

\citet{richards03} showed that the majority of UV-selected type 1 AGN at redshift $0.3<z<2.2$ exhibit an optical-UV continuum which, at wavelengths $\sim 1500-4000\AA$, is consistent with a single power-law. The optical-UV spectral slopes have a Gaussian distribution with $\aopt \sim -0.5 \pm 0.25$, where $\aopt$ is defined such that $L_\nu \propto \nu^\aopt$. However, the distribution of slopes also has a `red-tail', which includes AGN with redder continua than the general population, which also exhibit the characteristic continuum curvature expected from extinction by intervening dust. This sub-class, which includes $\sim 10\%$ of UV-selected AGN, is known as dust-reddened, or simply `reddened' QSOs. 

What is the value of $\aopt$ expected from models of geometrically thin, optically thick accretion disks (ADs)?
Such models are based on the general model presented in \citet{lyndenbell69} and \citet{shakura73} with various improvements like general relativistic corrections and radiative transfer in the disk atmosphere (e.g. \citealt{hubeny00}; \citealt{davis11}; \citealt{slone12}; \citealt{capellupo15}; \citealt{capellupo16}). Standard thin AD models are characterized by a slope of $\aopt\sim1/3$ at long wavelength. At intermediate frequencies which value depends on BH mass, BH spin and accretion rate, the slope becomes very flat with $\aopt\sim0$, while at the highest frequencies close to the peak temperature of the disk the slope becomes `red' with $\aopt<0$ \citep{pringle72}. For the typical mass and accretion rate of the sources considered in the present work ($L/\Ledd\sim0.1$, $\mbh\sim10^{8}\msun$, see below), the $3000-5100\AA$ slope predicted by the AD model is in the range $-0.1$ to $+0.2$, with some dependence on $\mbh$ and little dependence BH spin. This predicted slope is considerably bluer (`harder') than the typical observed slope of $\aopt\sim-0.5$ noted above. 
This discrepancy between the predicted and observed slopes has been pointed out by several studies which compared AD models with observed AGN spectra (\citealt{koratkar99}; \citealt*{davis07}; \citealt{jin12}; \citealt{shankar16}). More detailed work on a small sample (17 objects, see \citealt{shang05}) and a comprehensive work on a medium size sample (39 objects, see \citealt{capellupo15, capellupo16}) show better agreement with thin accretion disk spectra, especially when intrinsic reddening and the host contribution (at longer wavelength) is taken into account. 

In this work we explore the possibility that the discrepancy between predicted and observed slopes is due to extinction of the AD emission by dust along the line of sight (e.g.\ \citealt{netzer85}; \citealt{deZotti85}; \citealt{ward87}; \citealt{netzer95}; \citealt{bechtold97}; \citealt{gaskell04}; \citealt{richards06}; \citealt{davis07}; \citealt{dong08}; \citealt{stern12}; \citealt{dunn15}). That is, many AGN could also be extincted by dust, though by a smaller amount than the `red-tail' AGN mentioned above. 
Demonstrating that the majority of type 1 AGN are indeed mildly reddened would provide valuable observational support for the standard accretion disk models. 

Stern \& Laor (2012, hereafter SL12) provided evidence, based on a large number of Sloan Digital Sky Survey (SDSS) type 1 AGN, which supports the ubiquitous reddening scenario. They showed that the ratio of ultraviolet (UV) to broad H$\alpha$ emission correlates with $\aopt$ in luminous type 1 AGN, even when excluding the `red-tail' population. The slope of this correlation is consistent with the slope expected if the AD and broad line region (BLR) of typical type 1 AGN are extincted by dust. 
Also, the bluest $\aopt$ in the SL12 sample is consistent with AD predictions. 
These findings followed earlier studies based on smaller samples, which showed that the continuum slope is correlated with certain line ratios such as L$\alpha$/H$\beta$ (e.g. \citealt{netzer79, puetter81, soifer81, allen82, netzer95, bechtold97, baskin05}), where \cite{baskin05} also suggest that the dust has a planar distribution.
The implied reddening of a typical quasar with $\aopt\sim -0.5$ is $\ebv\approx0.07\mag$ (SL12), compared to significantly lower reddening values implied by assuming that the typical quasar is not reddened (\citealt{ richards03,hopkins04,lusso13,krawczyk15}).

Another prediction of the reddening scenario is that absorption features from gas associated with the dust should become stronger with decreasing $\aopt$. Testing this prediction is the main goal of the current paper.
While the relation between absorption features and spectral slope has been addressed by numerous papers (\citealt{sprayberry92}; \citealt{yamamoto99};  \citealt{hall02}; \citealt{reichard03}; \citealt{richards03}; \citealt{baskin05}; \citealt{shen12}; \citealt*{baskin13}; \citealt{dunn15}), these studies focused on absorption-selected samples. Since broad and narrow absorption lines are detected in only $\sim10\%$ and $\sim 2\%$ of UV-selected AGN, respectively (\citealt{richards03,shen12}), absorption-selected AGN are not necessarily representative of the general AGN population, which is the focus of the present work.

In this paper we wish to check whether the strength of absorption features correlates with spectral slope in the general AGN population, by comparing the spectral slope with the strength of the NaID absorption doublet ($\lambda\lambda5896,\, 5890$) in {\it stacked} SDSS spectra. Such a relation between NaID absorption and dust reddening in stacked spectra has been found in the MW and other galaxies. \citet*{poznanski12} used all the extragalactic objects from the SDSS to study the mean properties of the NaID absorption doublet in the MW and showed that there is a simple relation between the equivalent width (EW) of the doublet and dust reddening (see also \citealt{munari97, shih10}). There is also an ongoing effort to map the strength of the doublet in neighbouring galaxies, such as the small and large Magellanic clouds (SMC, LMC), which present different radiation fields and different gas-to-dust ratios from the MW (see for example \citealt{cox06b, welty06, van-loon13}, Welty et al.\ in prep.). 
The present paper extends this effort to AGN hosts.  

Our paper is organized as follows. 
In section \ref{s:data} we present the AGN sample, and measure the strength of the NaID absorption feature in stacked spectra as a function of continuum slope. We demonstrate that NaID absorption increases with decreasing continuum slope. 
In section \ref{sec: additional evidence} we provide additional evidence for the connection between continuum slope and dust reddening suggested by the observed EW(NaID) vs.\ continuum slope relation. We discuss our results in section \ref{sec: discussion}, and conclude in section \ref{s:conc}.

\section{Data and methods}\label{s:data}

\subsection{AGN sample}\label{s:sample}

\begin{figure}
\includegraphics[width=3.25in]{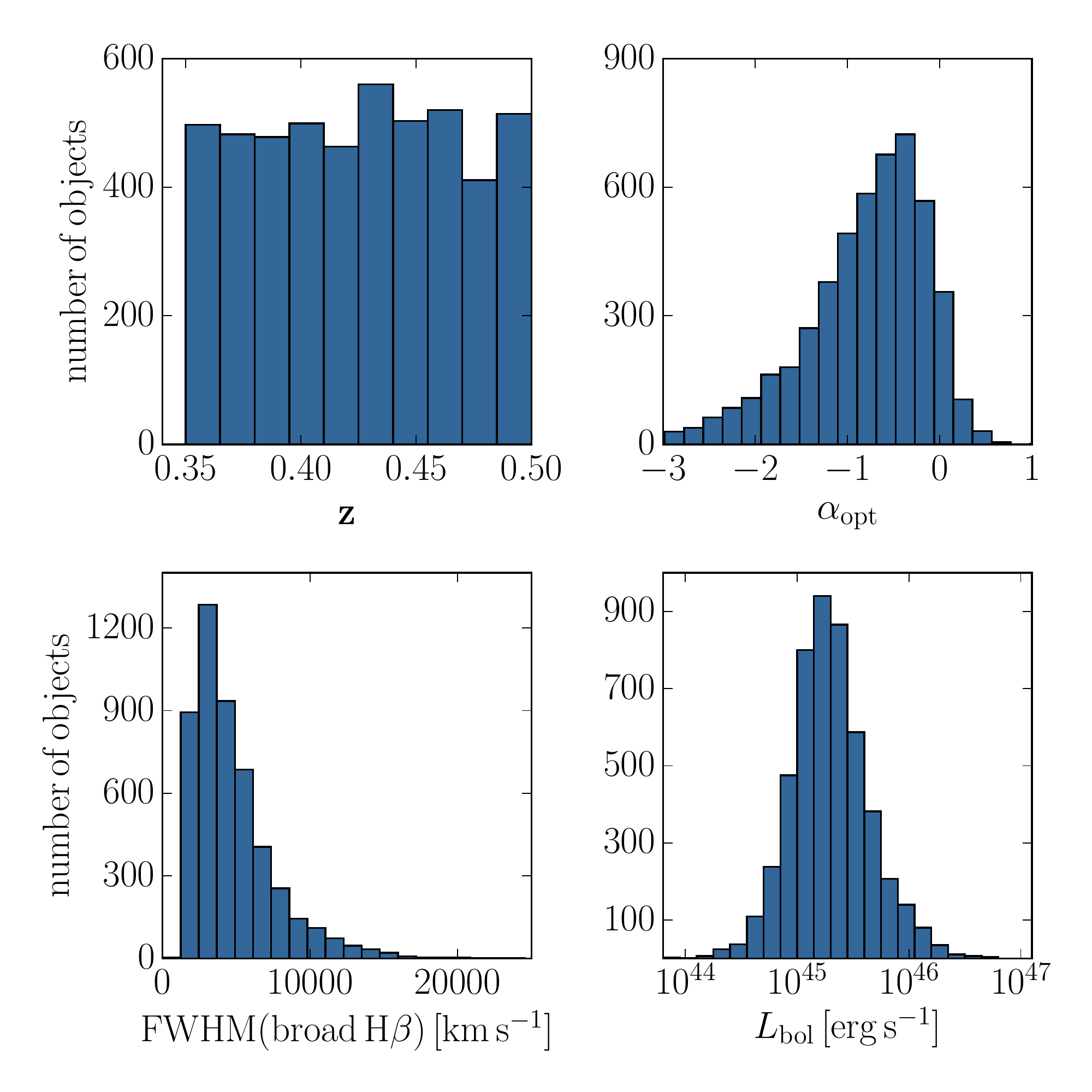}
\caption{
Histograms of different parameters of the AGN in our sample.
Redshift (top-left panel), 
continuum slope between 3000\AA\ and 5100\AA\ (top-right panel), 
the FWHM of the broad $\mathrm{H}\beta$ line (bottom-left panel), 
and the bolometric luminosity (bottom-right panel).
}\label{f:par_hist}
\end{figure}

The seventh data release (DR7) of the SDSS contains about a million extragalactic spectra covering the wavelength range of 3800 -- 9200\AA\ with 2.5\AA\ resolution. This includes the spectra of $105\,783$ AGN brighter than $M_{i}\,=\,-22.0\, \mathrm{mag}$ \citep{schneider10}, which have an average signal-to-noise ratio (SNR) of 10 pixel$^{-1}$. \citet{shen11} present a compilation of the properties for these AGN, including continuum and emission line measurements around the $\mathrm{H}\alpha$, $\mathrm{H}\beta$, and MgII wavelength regions which we use in this paper. 
The SDSS pipeline measures redshift, $z$, using the broad emission lines in AGN with errors of $\Delta z = 0.002$ (\citealt{hewett10}) compared to the host galaxy. 
We choose to use the redshifts measured by \citet{hewett10} based on the narrow emission lines, which are likely more accurate. 
We limit our sample to AGN with $0.35 < z < 0.5$ in order to include the NaID absorption doublet (5889.95\AA, 5895.92\AA) and the 3000\AA\, continuum in our wavelength window. 
The spectra are corrected for MW foreground dust, using the maps of \citet*{schlegel98} and the extinction law derived by \citet*{cardelli89}. 

Our final sample consists of $4\,946$ QSOs. 
We present in Figure~\ref{f:par_hist} their distribution in $z$, full width at half maximum (FWHM) of the broad $\mathrm{H}\beta$ line, and bolometric luminosity $\lbol$. The value of $\lbol$ is derived from the $L_{3000\AA}$ measurements of \cite{shen11} using the bolometric correction factor of \cite{richards06}. Fig.~\ref{f:par_hist} also shows the distribution of optical continuum slope $\aopt$, which is calculated as follows:
\begin{equation}\label{eq:1}
	{\aopt = \frac {\mathrm{log}[L_{\nu} (3000\mathrm{\AA}) / L_{\nu} (5100\mathrm{\AA})]} {\mathrm{log}(5100/3000)}} - 0.25
\end{equation}
where $L_{\nu} (\lambda)$ is the monochromatic luminosity at rest-frame wavelength $\lambda$ measured by \cite{shen11}. The offset of $-0.25$ is introduced since on average 13\%\ of the emission at 3000\AA\ is due to line and bound-free emission (\citealt{trakhtenbrot12}). 

The relation between $\aopt$ and the intrinsic AD slope depends both on the amount of extinction along the line of sight and on the host contribution to the continuum emission $\fhost(\lambda)$. We note that since we use a relatively small baseline to measure the slope ($\log (5100/3000)=0.23$), the relation between intrinsic emission and $\aopt$ is quite sensitive to $\fhost$ at $5100\AA$ and $3000\AA$, e.g.\ a value of  $\fhost=0.1$ at either wavelength will change the implied intrinsic slope by $\pm0.2$. We discuss the effect of $\fhost$ further below.

We interpolate each spectrum to an identical grid of 0.5\AA\ in rest-frame wavelength, in the range 5700 -- 6100\AA, which contains the NaID absorption doublet. We use the Savitzky-Golay (SG) smoothing algorithm \citep{savitzky64}\footnote{SG is a generalization of `running mean' where instead of fitting a constant in the window (the mean) one can fit a polynomial.} with a 3rd order polynomial fit and a moving window size of $35$\,\AA\ to fit and divide out the continuum, thus obtaining normalized fluxes. We exclude from the fitting range the wavelengths 5875 -- 5920\AA, in order to avoid the HeI emission line (5875.6\AA) and NaID absorption lines (5889.95\AA, 5895.92\AA). The NaID absorption line is not detected in almost all individual spectra, due to the limited SNR of the SDSS spectra. 

\begin{figure*}[!ht]
	\center
\includegraphics[width=0.9\textwidth]{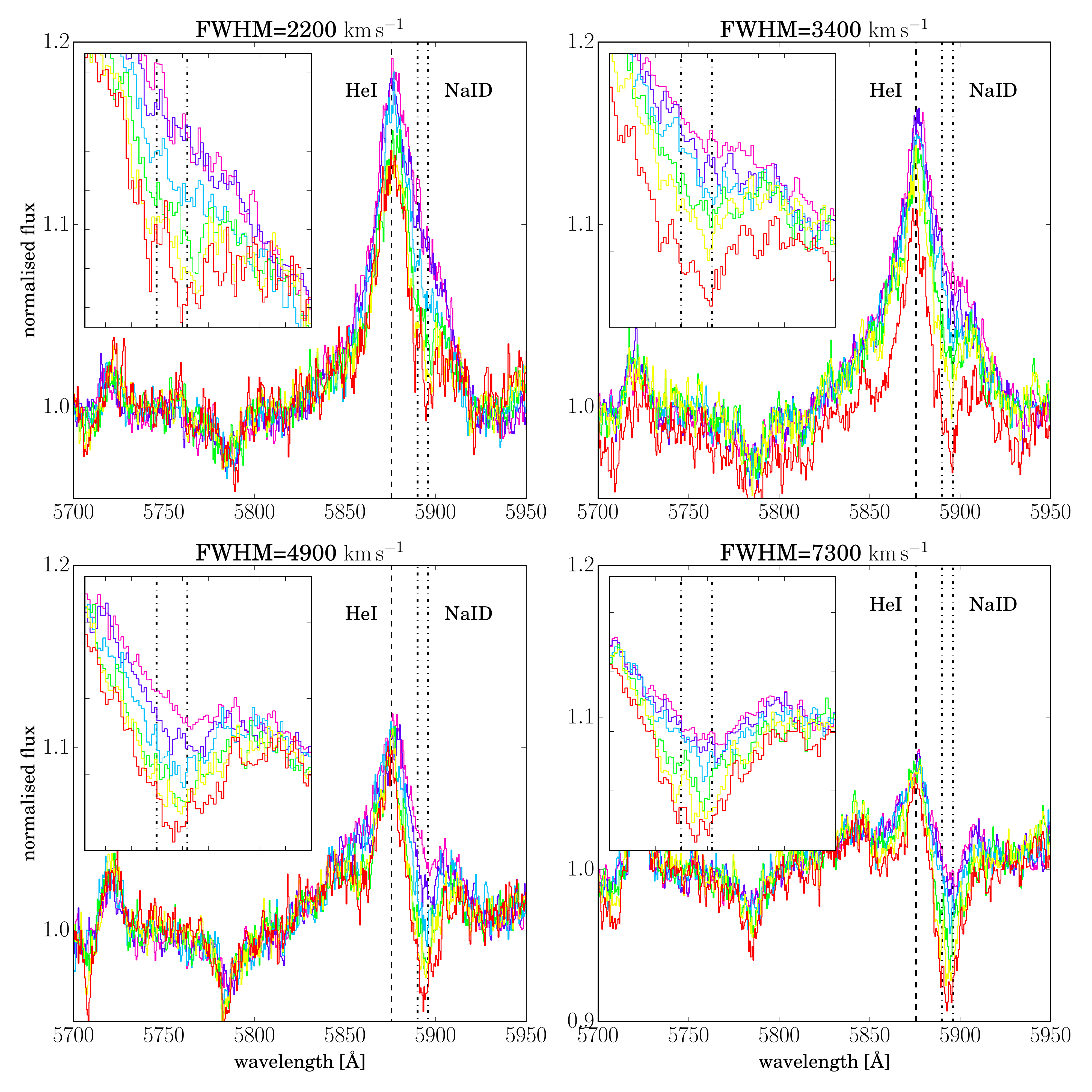}
\includegraphics[width=0.6\textwidth]{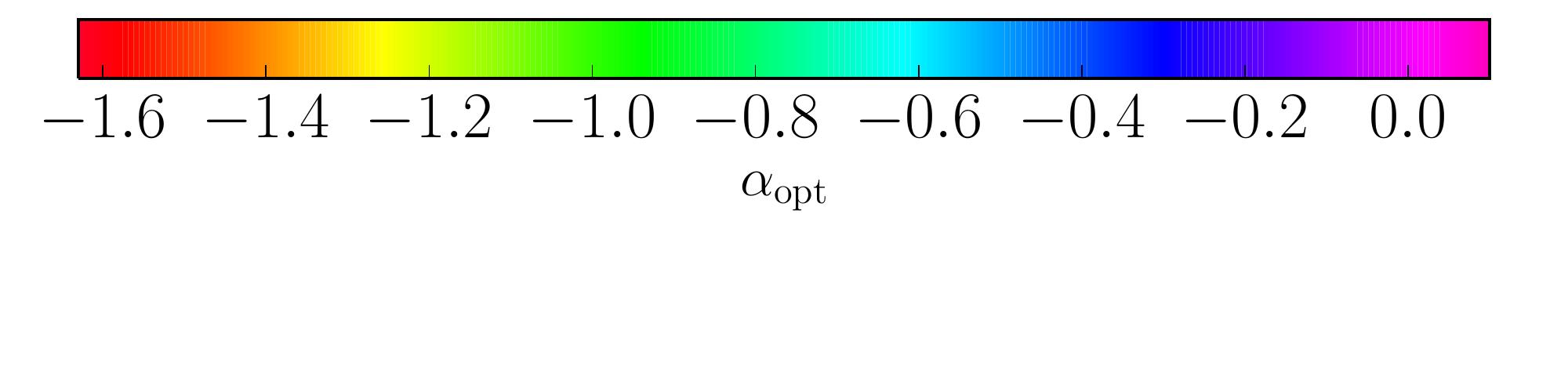}
\caption{Stacked AGN spectra at wavelengths near the NaID absorption features, in bins of FWHM(H$\beta$) and continuum slope. The spectra are normalized by the continuum level using a Savitzky-Golay smoothing algorithm. 
The insets zoom in on the NaID doublet. Note how EW(NaID) grows with decreasing $\aopt$.
} \label{f:stacked_example}
\end{figure*}

\subsection{Stacked spectra}\label{s:method}

We divide the spectra into two-dimensional bins based on their optical slope and the FWHM of their broad $\mathrm{H}\beta$ line. 
The latter binning is done since the NaID absorption doublet lies on the red wing of the broad HeI feature, which has a width that correlates tightly with the broad H$\beta$ width. Therefore, stacking in bins of FWHM avoids spurious features which are created when stacking objects with very different FWHM. 
We find that in order to reach a sufficient SNR to measure the depth of the NaID line, we must stack at least 200 spectra. In order to have a similar number of spectra in every bin, we choose the following FWHM bins: $1000-3000$, $3000-4000$, $4000-6000$, and $6000-10000\kms$. 
Each of the FWHM bins contains roughly 1200 spectra, and we divide them into six $\aopt$ bins, which we stack using the median at each wavelength. Our final sample contains 24 stacked spectra with SNR in the range $27-39$.

\begin{figure*}[t]
	\center
\includegraphics[width=0.9\textwidth]{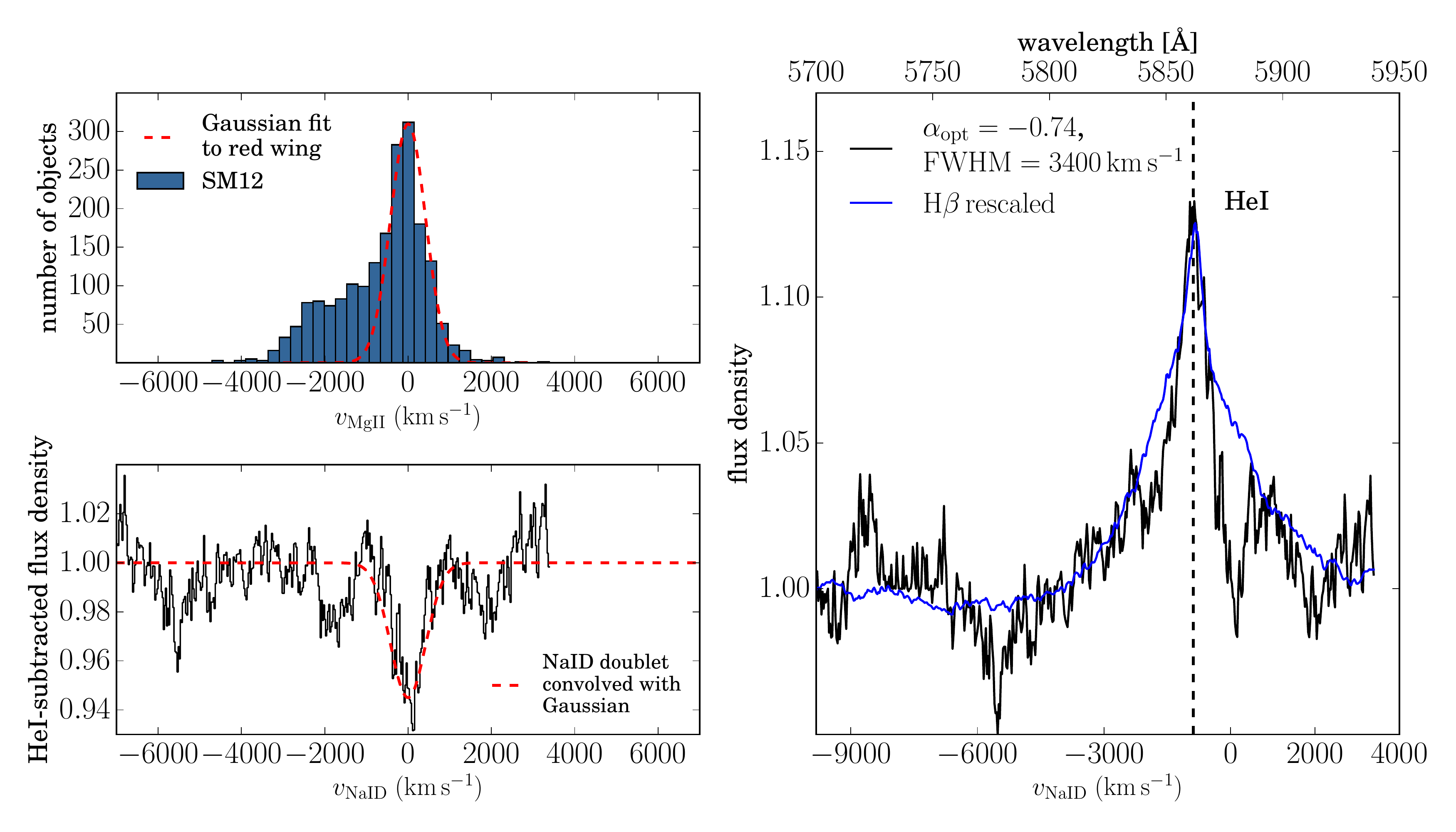}
\caption{
{\bf (Top-left)} 
The velocity distribution of associated MgII absorbers from SM12 (blue bars). The dashed line marks a Gaussian fit to the red wing of the distribution. 
{\bf (Right)} 
A comparison of the HeI and H$\beta$ profiles in the stacked spectrum of AGN with $\aopt=-0.74$ and ${\rm FWHM}=3400\kms$. We use the rescaled H$\beta$ as a template for HeI. {\bf (Bottom-left)} A comparison of the stacked spectrum of AGN with $\aopt=-0.74$ and ${\rm FWHM}=3400\kms$, after subtracting out the HeI feature (black solid line), with the profile of a NaID doublet convolved with the Gaussian velocity distribution of the MgII absorbers (red dashed line).
The velocity distribution of the MgII absorbers is consistent with the profile of the stacked NaID absorption feature. 
}
\label{f:voff_dist_comb_prof}
\end{figure*}

Figure \ref{f:stacked_example} shows the six stacked spectra with different optical slopes for each FWHM bin. These are the independent spectra with which we perform the fitting throughout the paper. One can see that the NaID absorption profile lies on the red wing of the HeI emission line at 5876\,\AA. 
Although there is some variation in the HeI profile between different $\aopt$ bins, the EW of the NaID absorption grows markedly as $\aopt$ decreases. 

One can see additional absorption features in the stacked spectra. The absorption line at 5780\AA\, shows little change as a function of the optical slope and its equivalent width is approximately 2\,\AA\, throughout the different bins. We speculate that this line is due to FeI absorption (5780\,\AA). However, a certain identification requires a detailed comparison with the strengths of other known Fe absorption lines (in the 4000--5600\,\AA\, wavelength range) and is beyond the scope of our paper. Figure \ref{f:stacked_example} show an additional absorption feature at 5855\AA, on the blue side of the HeI emission line, which we call the blue absorption feature (BAF). The origin of the BAF is likely not NaID, since it is usually a resolved trough and is offset from the rest wavelength of NaID by about $1800\,\mathrm{km}\,\mathrm{s}^{-1}$. The BAF strength is not correlated with the strength of other photospheric features strength (such as FeI 5780\,\AA). However, we account for this absorption when modelling the HeI and NaID lines.

In most bins the NaID doublet appears as a single blended profile, suggesting the width of the individual lines are typically larger than their wavelength separation of $6\AA$ ($300\kms$). Only in the reddest spectra of the FWHM$=2200\kms$ bin is the doublet clearly resolved. 
The combined profile of the NaID doublet has a FWHM in the range of 7--9\AA\ ($350-450\kms$).

\begin{figure*}
	\center
\includegraphics[width=0.85\textwidth]{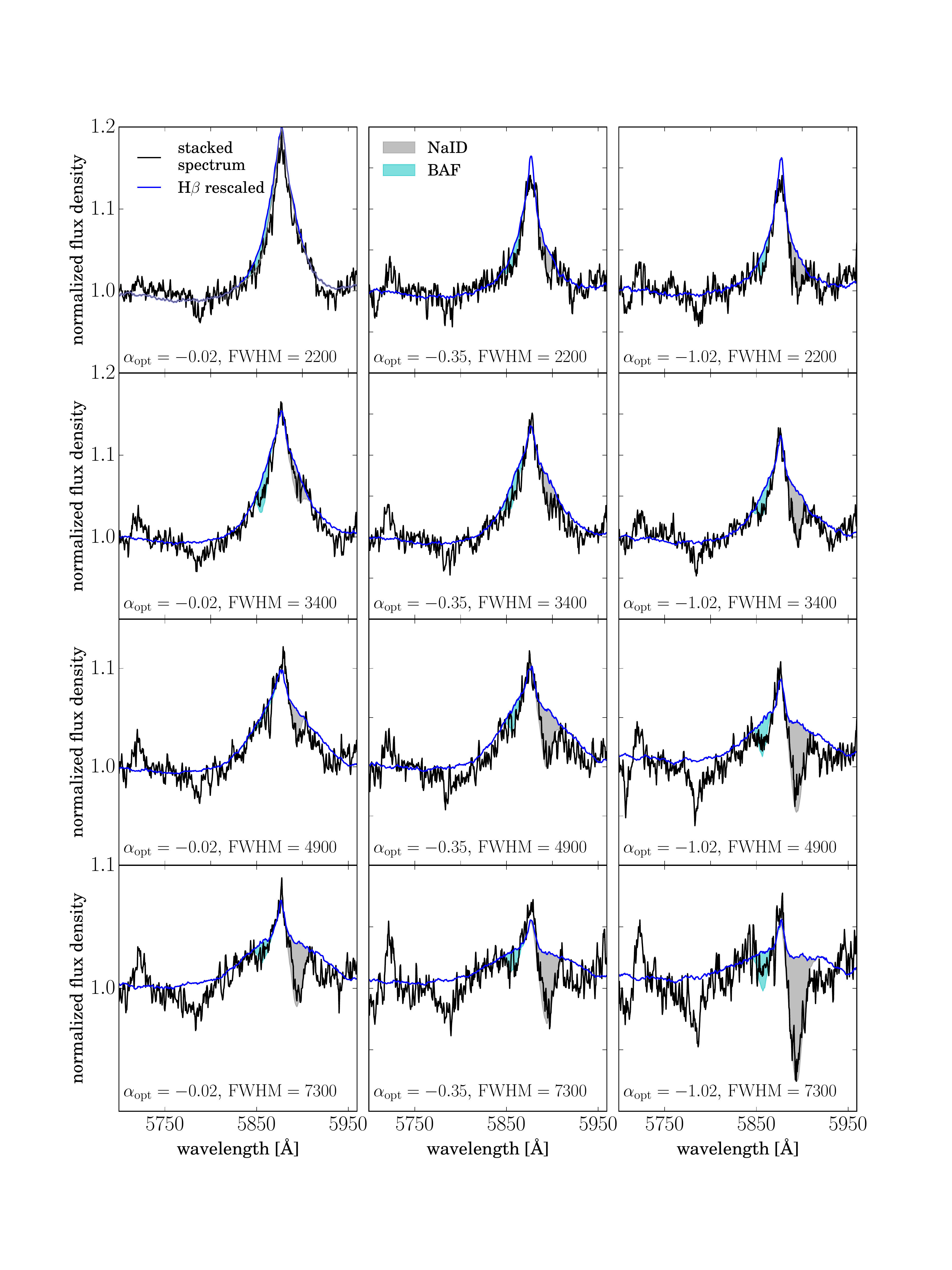}
\caption{
Fits to the different components around the NaID absorption feature.
The different rows represent different FWHM bins while the different columns represent different $\aopt$ bins.
Black lines plot the normalized stacked spectra, while blue lines plot the shifted and rescaled $\mathrm{H}\beta$ of the same spectrum, which we use as a template for HeI. The fit NaID absorption feature is colored in gray, while the fit to the unexplained absorption feature on the blue side of HeI is colored in cyan.
 }\label{f:Hb_He_example}
\end{figure*}

In Figure~\ref{f:voff_dist_comb_prof}, we compare the velocity profile of the stacked NaID absorption feature with the velocity distribution of narrow MgII absorbers found by \citet[][hereafter SM12]{shen12}. SM12 found that $2\%$ of SDSS AGN at $0.4 < z < 2$ exhibit a narrow MgII absorption feature in single spectra. The top-left panel in Fig.~\ref{f:voff_dist_comb_prof} shows the velocity distribution of the SM12 absorbers. As in this study, the systemic velocity of the host in the MgII-absorber sample is derived from the velocity of the narrow emission lines. 
SM12 argue that MgII absorbers with velocities $v < -1500\kms$ 
are consistent with cosmologically intervening absorbers along the line of sight towards the AGN (see also \citealt{wild08}) since the velocity distribution profile flattens out at $v < -1500\kms$. 
Therefore, to assess the velocity distribution of the associated (non-intervening) absorbers, we fit the positive velocity part of the distribution with a Gaussian. 
In the bottom-left panel we compare this MgII velocity distribution with the NaID absorption profile of the $\aopt=-0.74,~{\rm FWHM}=3400\kms$ bin. 
This bin is chosen since it shows strong NaID absorption which facilitates the comparison.
The black line is the stacked AGN spectrum after subtracting out the HeI line, where we use the H$\beta$ profile as a template for HeI. The profiles of HeI and H$\beta$ are compared in the right panel. The red dashed line in the bottom-left panel is the expected profile for a NaID doublet convolved with the Gaussian shown in the top-left panel. 
Fig.~\ref{f:voff_dist_comb_prof} shows that the velocity profile of the stacked NaID absorption feature is consistent with the velocity distribution of associated MgII absorbers found by SM12. 

\subsection{Equivalent width measurements}\label{s:naid_ew}

In order to measure the EW of the NaID line in the stacked spectra, we perform a joined fit. The HeI line is modeled with the H$\beta$ profile of the same stacked spectrum. 
The NaID doublet profile is modeled as a single unresolved line with the width suggested by the MgII-absorption distribution shown in Fig.~\ref{f:voff_dist_comb_prof}. The BAF strength correlates with $\aopt$, and therefore changes from one stacked spectrum to the next. We fit a Gaussian function to account for it. To solve for the various parameters of interest, we minimize the $\chi^{2}$ of the following function:
\begin{equation}\label{eq:2}
	{f(\lambda) = a_{H}\cdot \mathrm{H}{\beta}(\lambda) + b  + a_{N}\cdot N(\lambda) + G(a_G, \sigma, \lambda_{0},\lambda)}
\end{equation}
where $\mathrm{H}{\beta}$ is the shifted and rescaled $\mathrm{H}{\beta}$ profile (with a fixed width), $N$ is the NaID profile, and $G$ is the Gaussian function that accounts for the BAF. We allow the continuum parameter, $b$, the scaling parameters $a_H$, $a_N$, and $a_G$, and the width of the BAF, $\sigma$, to change from one stacked spectrum to the next. The central wavelengths for all the features are fixed. The NaID EW is then extracted from the best fit. Different fitting procedures, e.g., using a Voigt profile for the BAF and Gaussians for the NaID doublet, give essentially the same EWs. Figure \ref{f:Hb_He_example} shows the quality of the fit we obtain for different FWHM and $\aopt$ bins. The stacked spectra are shown in black, the HeI fits in blue, and the NaID and BAF features are shaded in gray and cyan respectively. 

\subsection{EW vs.\ $\aopt$}\label{s:results}

\begin{figure}
\includegraphics[width=3.25in]{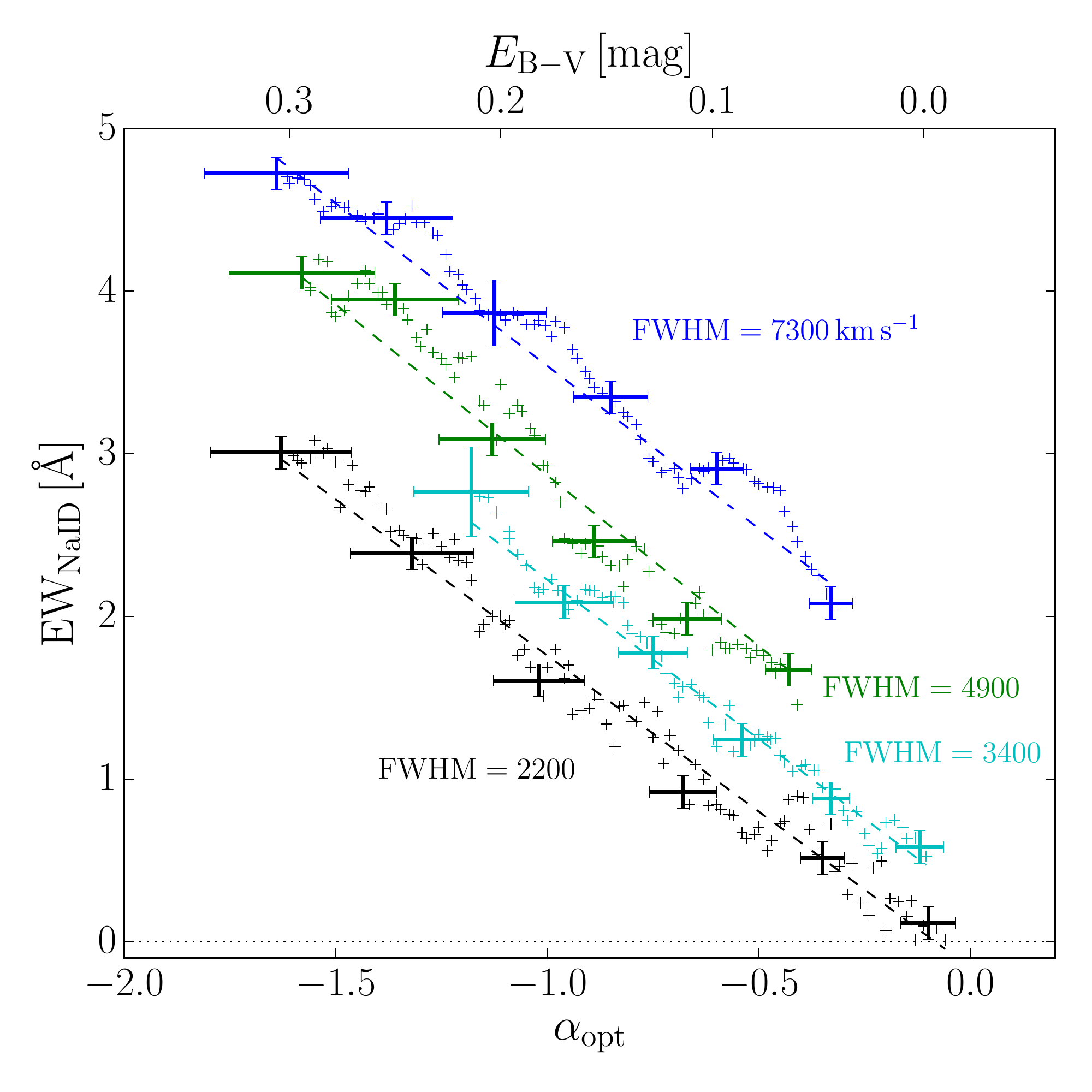}[!h]
\caption{
The relation between EW(NaID) and optical slope in stacked AGN spectra with different FWHM(H$\beta$). 
The large crosses mark the median EW and $\aopt$ measured on independent stacked spectra, while the small crosses mark the running medians. 
The fitted linear relations are plotted as dashed lines and listed in Table~\ref{tab: relations}. The slopes of the relations are similar in the different FWHM bins. 
In the ${\rm FWHM}=2200\kms$ bin, we find ${\rm EW(NaID)}=0$ at $\aopt=-0.1$. This $\aopt$ is similar to the intrinsic continuum slope expected from accretion disk with BH mass and accretion rate similar to the ones in our sample. 
We therefore suggest that most type 1 AGN are reddened by dusty-gas along the line of sight, where the dust reddens the observed slopes to $\aopt < -0.1$. 
Only in the few AGN with $\aopt \approx -0.1$ the intrinsic accretion disk spectrum is not extincted. 
The implied $\ebv$ for each $\aopt$ (eqn.~\ref{eq: aopt vs ebv}) are noted on top.  }
\label{f:ew_alpha_all}
\end{figure}

For a given FWHM bin, in addition to dividing the spectra into six bins of different $\aopt$, we sort the spectra in $\aopt$ and use a running median filter with a precision of $0.01$ in $\aopt$. Using the running median, we essentially incorporate the information of every low SNR spectrum into many high SNR spectra. The stacked spectra of the running median are therefore not statistically independent. We measure the EW of these spectra and present them as small crosses (with no uncertainties) in Figure \ref{f:ew_alpha_all}. This is done in order to trace the behaviour qualitatively, and examine the scatter around each of the 24 independent stacked spectra.

Additionally, we measure the EW of NaID and the median $\aopt$ for 24 independent spectra. The uncertainties in $\aopt$ are dominated by the uncertainty of $0.08$ in the correction for emission lines discussed in section \ref{s:sample}. The uncertainties in EW(NaID) are calculated during the line fitting. 

\begin{deluxetable}{cccc}
\tablewidth{0pt}
\tabletypesize{\scriptsize}
\tablecaption{EW(NaID) and $\aopt$ relations\label{tab: relations}}
\tablehead{
\colhead{FWHM $[\kms]$} &
\colhead{$\alpha_0$} &
\colhead{EW at $\alpha_0$ [\AA]} &
\colhead{EW vs.\ $\aopt$ slope}}
\startdata
2200 & -0.11 & $0.05 \pm 0.10$ & $-1.92 \pm 0.09$ \\
3400 & -0.12 & $0.58 \pm 0.14$ & $-1.96 \pm 0.08$ \\
4900 & -0.43 & $1.67 \pm 0.16$ & $-2.10 \pm 0.10$ \\
7300 & -0.33 & $2.08 \pm 0.19$ & $-2.00 \pm 0.11$ \\
\enddata 
\tablecomments{Fit parameters for the linear relations between EW(NaID) and $\aopt$ shown in Fig.~\ref{f:ew_alpha_all}.}
\end{deluxetable}

Figure~\ref{f:ew_alpha_all} shows the EW(NaID) vs.\ $\aopt$ relation for the four FWHM bins. 
The measured EW(NaID) is linearly proportional to $\aopt$, with similar slopes in the different FWHM bins. 
We fit a linear function of the form
\begin{equation}
 {\rm EW}({\rm NaID}) = a\left(\alpha-\alpha_0\right) + b
\end{equation}
to the six independent measurements of EW versus $\aopt$ in each FWHM bin, where $\alpha_0$ is the bluest $\aopt$ at each FWHM. 
We use an Orthogonal Least Square (OLS) minimization, which takes into account the uncertainties in both variables. 
The best linear fits are shown in Fig.~\ref{f:ew_alpha_all} and listed in Table~\ref{tab: relations}.

\section{Additional evidence for the connection between $\aopt$ and $\ebv$}\label{sec: additional evidence}

What is the underlying physical explanation of the EW(NaID) vs.\ $\aopt$ relations seen in Fig.~\ref{f:ew_alpha_all} and listed in Table~\ref{tab: relations}?
In observed AGN spectra, $\aopt$ is expected to depend on three components: the slope of the disk SED, the slope of the stellar SED, and the reddening of both.
The NaID absorption feature can originate from gas along the line of sight to the AGN or to the stars, but also from absorption in the photospheres of the stars. 

Fig.~\ref{f:ew_alpha_all} shows that in the lowest-FWHM bin, EW(NaID) goes to zero at $\aopt = -0.1$. 
As explained in section \ref{s:intro}, and discussed some more in section \ref{s: coherent picture} below, this value of $\aopt$ is similar to the intrinsic spectral slope expected in accretion disks around BHs in the mass and accretion rate range considered here after allowing for host galaxy contribution. 
Hence, objects which show no NaID absorption also exhibit the continuum slope expected from accretion disk models. 
This correspondence is consistent with a scenario where the observed distribution in $\aopt$ is driven by the distribution of column of dusty-gas along the line of sight, as previously suggested by \cite{ward87}, \cite{gaskell04} and SL12.
The blue intrinsic AGN SED is seen only in the few objects with $\aopt=-0.1$ and EW(NaID)$\,\sim0$, while most objects show a redder $\aopt\approx-0.5$ and ${\rm EW(NaID)} \approx 0.8\AA$, due to extinction by dust and absorption by gas associated with the dust. 
In the three large-FWHM bins, the slopes of the EW vs.\ $\aopt$ relations  are similar to the slope in the FWHM$=2200\kms$ bin. 
This similarity suggests that the EW vs.\ $\aopt$ relations in all FWHM bins could be driven by the distribution of dusty-gas columns along the line of sight. In this section we provide additional observational evidence supporting this conjecture, and in the discussion we confront this conjecture with photoionization calculations of gas exposed to AGN radiation.

\begin{figure}
\includegraphics[width=0.5\textwidth]{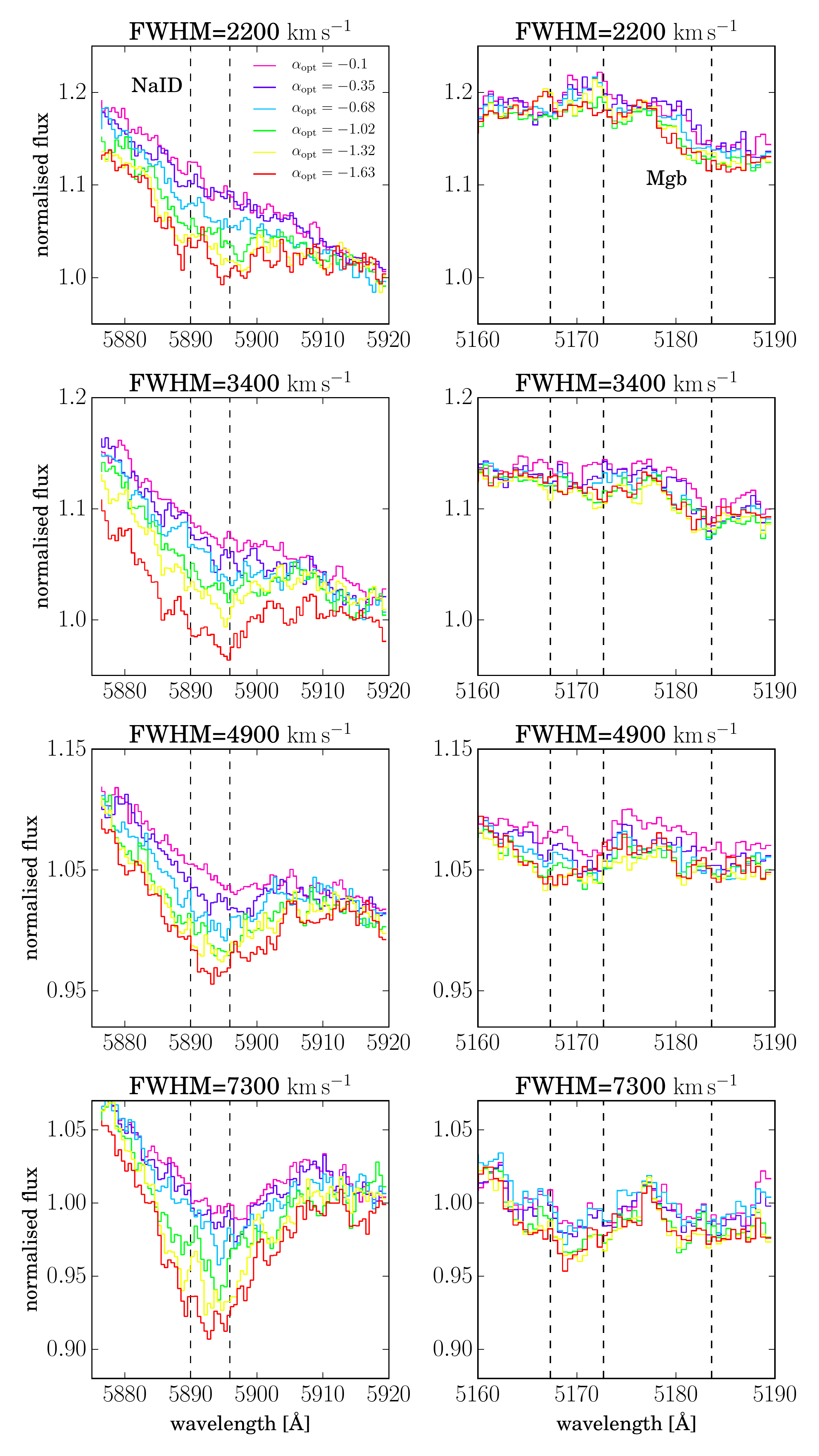}
\caption{Comparison of NaID absorption (left panels) to the Mg\,b absorption complex (right panels) in the stacked and normalized spectra shown in Fig.~\ref{f:stacked_example}. Different rows plot different FWHM bins and different colors denote different $\aopt$ bins. Mg\,b absorption does not originate in the ISM, and hence it provides an estimate of the contribution of stellar photospheres to the trends of EW(NaID) vs.\ $\aopt$ and EW(NaID) vs.\ FWHM seen in Fig.~\ref{f:ew_alpha_all}.
For a given FWHM($\mathrm{H\beta}$) bin, Mg\,b absorption remains roughly constant as a function of $\aopt$, in contrast with the significant increase in NaID absorption with decreasing $\aopt$. Both Mg\,b and NaID absorption become stronger with increasing FWHM. 
This figure reinforces our interpretation that the trend of EW(NaID) with $\aopt$ is not associated with stellar photospheres.
}
\label{f:mg_stacks}
\end{figure}

\subsection{Contribution from stellar photospheres}

In Figure~\ref{f:mg_stacks} we compare the NaID absorption feature with the Mg\,b absorption complex. 
Mg\,b absorption likely originates in stellar photospheres, since it requires a significant population of excited ions, while in the ISM most ions are at the ground state. Hence, we can use Mg\,b to estimate the stellar photosphere contribution to the trends of EW(NaID) vs.\ $\aopt$ and  EW(NaID) vs.\ FWHM seen in Fig.~\ref{f:ew_alpha_all}.
The right panels in Fig.~\ref{f:mg_stacks} show the Mg\,b absorption features in the normalized spectra from Fig.~\ref{f:stacked_example}, where different rows plot different FWHM bins and different colors denote different $\aopt$ bins. 
These features can be compared to the NaID features shown in the left panels.

Measuring EW(Mg\,b) accurately is complicated by the fact that its wavelengths coincide with the wavelengths spanned by the optical BLR FeII forest. However, it is evident from Fig.~\ref{f:mg_stacks} that in a given FWHM bin, the Mg\,b feature does not  significantly change with $\aopt$, in contrast with the significant trend apparent in NaID. The lack of trend in Mg\,b with $\aopt$ suggests that the trend of NaID with $\aopt$ is not due to an increasing contribution from stellar photospheres. 
In contrast, the strength of the Mg\,b features increases with increasing FWHM. 
Therefore, it seems plausible that the trend of EW(NaID) with FWHM is associated with an increasing contribution from stellar photospheres. The trend of EW with FWHM is further addressed in the discussion. 

A stellar photosphere origin is also unlikely to explain the largest EW(NaID)$=4.8\AA$ we observe (Fig.~\ref{f:ew_alpha_all}) since the maximum EW(NaID) seen in the spectrum of individual stars is 5.6\AA\ (\citealt{burstein84, faber85}). Therefore, in order to explain an EW of 5\AA\ with stellar photospheres, one has to populate the entire AGN host with stars which have the maximum EW(NaID), and assume the contribution of the AGN continuum at 6000\AA\ is entirely negligible, which is unlikely. 

\subsection{H$\alpha$/H$\beta$ vs.\ $\aopt$}\label{s:balmer}

Various studies suggest that the BLR resides within the dust sublimation radius (\citealt{netzer93, kaspi00, suganuma06, bentz09, koshida14}),
 and hence any dust surrounding the AGN will reside on larger scales. 
 The BLR is therefore likely to also be extincted by the dust which reddens the AGN continuum. 
 In this section we test this prediction by comparing the observed BLR Balmer ratio $\Ha$/$\Hb$ with $\aopt$. 
 This is somewhat similar to \cite{dong08} where the aim was to find the line ratio in the bluest (largest $\aopt$) AGN. 
 As shown below, our analysis is more general and apply to the entire sample rather than the sub-group with the smallest $\Ha$/$\Hb$.

To calculate the expected $\Ha$/$\Hb$ as a function of $\aopt$ in our scenario, we quantify the dust extinction in magnitude at wavelength $\lambda$ as $R_\lambda\ebv$, where $\ebv$ is the color excess which is proportional to the dust column, and $R_\lambda$ is the assumed extinction law. 
Since in our suggested picture $\aopt$ is determined by the dust column along the line of sight, 
$\aopt$ is related to $\ebv$ via
\begin{equation}\label{eq:ebv_alpha}
\alpha_0 - \aopt  = \frac{R_{3000} - R_{5100}}{2.5\log(5100/3000)}\,\ebv ~.
\end{equation}
Assuming the reddening dust completely covers the BLR, then 
\begin{equation}\label{eq:balmer_dec_standard}
 \log\left( \frac{\Balmerint}{\Balmerobs}\right) = \frac{R_{4861} - R_{6563}}{2.5}\ebv ~,
\end{equation}
where $\Balmerint$ and $\Balmerobs$ are the intrinsic and observed H$\alpha$/H$\beta$ ratios, respectively. 
Combining equations~(\ref{eq:ebv_alpha}) and (\ref{eq:balmer_dec_standard}) we therefore expect
\begin{eqnarray}\label{eq:dec_alpha}
\log(\frac{\Balmerobs}{\Balmerint} &=& 0.23\,(\alpha_0 - \aopt)\frac{R_{4861} - R_{6563}}{R_{3000} - R_{5100}} \nonumber \\
&\approx& 0.1 (\alpha_0 - \aopt) ~.
\end{eqnarray}
The approximation in eqn.~(\ref{eq:dec_alpha}) is based on the \cite{pei92} formulation for the MW, LMC, and SMC-type extinction curves, in which $0.23(R_{4861} - R_{6563})/(R_{3000} - R_{5100})$ is equal to $0.099$, $0.100$, and $0.096$, respectively. The similarity of these values is the result of the narrow wavelength range considered here. It suggests that the exact nature of the extinction law is a minor source of uncertainty. 
In low density gas $\Balmerint$ is equal to $2.74-2.86$, assuming `Case B' recombination (\citealt{osterbrock06}). 
However, in the dense BLR gas ($> 10^9\cm^{-3}$), the value of $\Balmerint$ may be different due to line optical depths and collisional effects \citep{netzer13}.

\begin{figure*}
	\center
\includegraphics[width=0.99\textwidth]{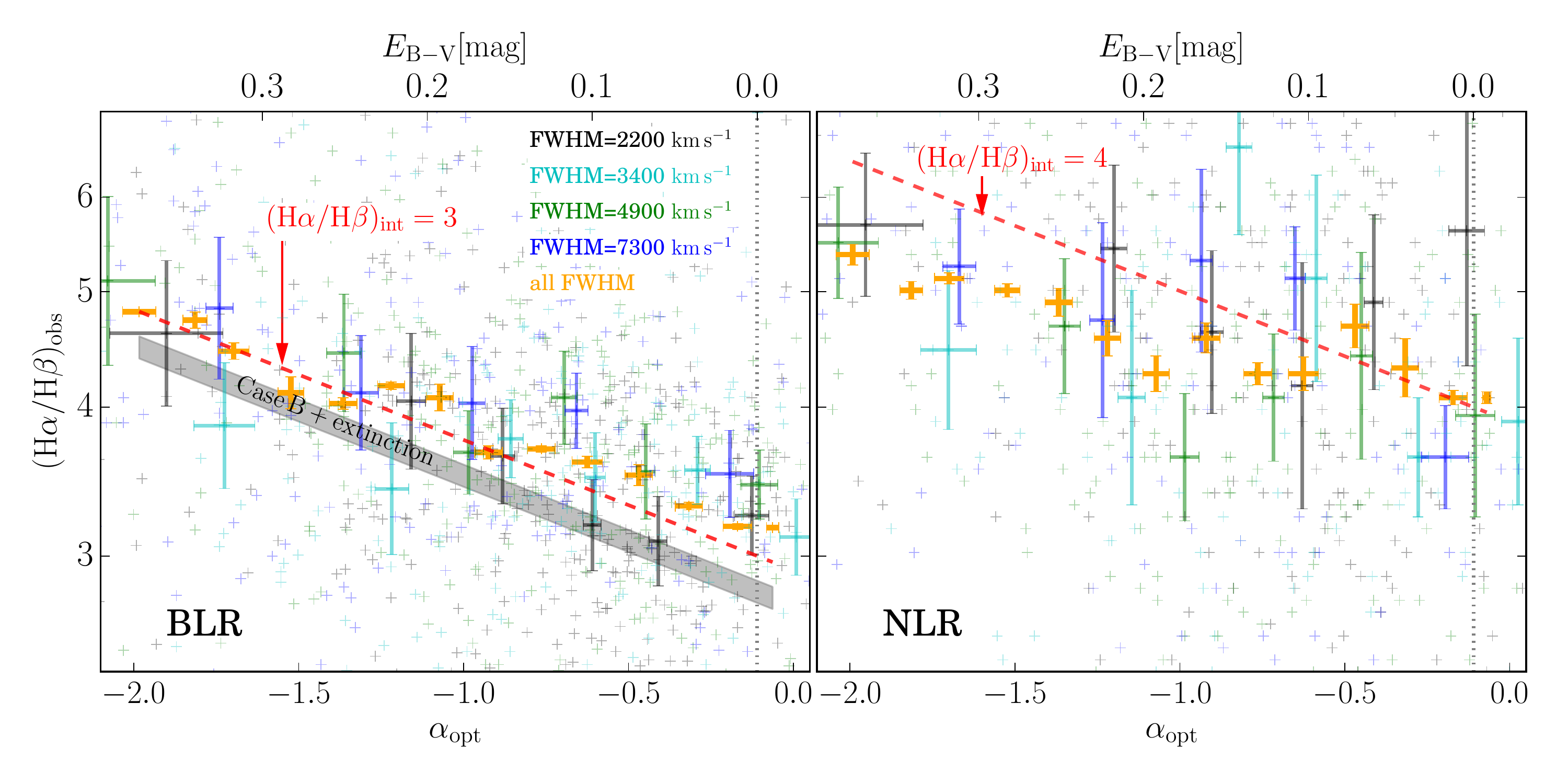}
\caption{Comparison of the observed {\rm H}$\alpha$/{\rm H}$\beta$ ratios in our sample with the expected values if $\aopt$ is determined by the dust column along the line of sight. 
{\bf (Left panel)} 
Orange error bars mark the mean observed {\rm H}$\alpha$/{\rm H}$\beta$ in the BLR vs.\ $\aopt$. The black, cyan, green, and blue error bars mark the observed {\rm H}$\alpha$/{\rm H}$\beta$ in different FWHM bins, while the crosses (of the same colours) represent individual measurements.
The implied $\ebv$ for each $\aopt$ are noted on top.
Red dashed line marks the expected observed {\rm H}$\alpha$/{\rm H}$\beta$ vs.\ $\ebv$ relation for an assumed intrinsic Balmer ratio of 3.
The observed and expected Balmer ratios match for $\Balmerint=3.02$, close to the case B value, supporting our suggestion that $\aopt$ is set by the dust column along the line of sight. 
{\bf (Right panel)}
A similar comparison for the NLR. In the NLR a $\Balmerint$ larger than the Case B value is expected due to dust extinction within the NLR clouds. The shallower slope of the NLR vs. $\Balmerobs$ compared to the expected slope may suggest that the reddening dust partially covers the NLR.
}
\label{f:blamer}
\end{figure*}

To derive the median $\Balmerobs$ as a function of $\aopt$, we use the catalog of \citet{shen11} who measured the broad components of the $\mathrm{H\alpha}$ and $\mathrm{H\beta}$ emission lines. As the $\mathrm{H\alpha}$ emission line is observable only for redshifts $z < 0.4$, we use only the 1296 AGN in the redshift range $0.35 < z < 0.4$. 
We group the AGN into bins with width $\Delta\aopt=0.1$, and plot the median $\Balmerobs$ in each bin in the left panel of Figure~\ref{f:blamer}.
The uncertainty on $\aopt$ is calculated in section \ref{s:results} 
and the uncertainty of the median $\Balmerobs$ is the weighted median of the uncertainties in every $\aopt$-bin, where the uncertainties in each bin are measured using the line measurement uncertainties given by \citet{shen11}. We also show the median $\Balmerobs$ in different FWHM bins, where each $\aopt$ bin contains the same number of objects.

The dashed lines in the left panel of Fig.~\ref{f:blamer} are the expected relations between $\Balmerobs$ and $\aopt$ (eqn.~\ref{eq:dec_alpha}), for $\alpha_0=-0.1$ and $\Balmerint=3$. 
We find a best-fit $\Balmerint=3.02\pm0.21$, close to the value expected from Case B recombination. 
This best-fit value is consistent with the Balmer ratio of blue AGN previously measured by \cite{dong08}. 
The uncertainty of 0.21 on the mean $\Ha/\Hb$ ratio is due to the uncertainties of the mean ratios and the best fit. 

It is important to note that the distribution of $\Ha/\Hb$ ratios for the bluest objects is quite broad ($3.02 \pm 1.19$), and there are many objects with $\Ha/\Hb$ significantly larger and significantly smaller than $3$ with individual uncertainties that clearly distinguish them from the mean. Thus, we see no evidence for a `single-value $\Ha/\Hb$', which is also supported by \citet{schnorr-muller16}, who study that BLR of nine Seyfert 1 galaxies using up to six broad HI lines in each spectrum.
Fig.~\ref{f:blamer} demonstrates that the slope of the relation between $\Balmerobs$ and $\aopt$ is consistent with the expected slope. 
The best fit slope of the entire sample is :
\begin{equation}\label{eq:balmer_dec_full_stack}
 \log\,\frac{\Balmerobs}{3.02 \pm 0.21} = (0.0950 \pm 0.053)\,(\alpha_0 - \aopt)
\end{equation}, 
while the best-fit for each of the FWHM 2200, 3400, 4900, and 7300 $\mathrm{km\,s^{-1}}$ bins is 
\begin{eqnarray}\label{eq:balmer_dec_fwhm_bins}
 \log\,\frac{\Balmerobs}{3.01 \pm 0.42} = \,\,\,(0.103 \pm 0.0091)\,(\alpha_0 - \aopt)  \nonumber \\
 \log\,\frac{\Balmerobs}{3.36 \pm 0.51} = \,\,\,\,\,\,(0.064 \pm 0.016)\,(\alpha_0 - \aopt)  \nonumber  \\
 \log\,\frac{\Balmerobs}{3.37 \pm 0.47} = (0.0876 \pm 0.0094)\,(\alpha_0 - \aopt)  \nonumber \\
 \log\,\frac{\Balmerobs}{3.50 \pm 0.53} = (0.0728 \pm 0.0079)\,(\alpha_0 - \aopt)
\end{eqnarray}
respectively. Therefore, the dependence of $\Balmerobs$ on $\aopt$ supports our interpretation that $\aopt$ is determined by the dust column along the line of sight, such that $\ebv$ can be derived from $\aopt$ via eqn.~(\ref{eq:ebv_alpha}).

Given the typical $\ebv$, we can also study the effect of reddening on additional broad lines at shorter wavelength, in particular the $\mathrm{L\alpha}$/$\mathrm{H\beta}$ ratio. For galactic type reddening, and $\ebv = 0.08\mag$, we would predict the observed ratio to be 3.8 times smaller than the intrinsic one. This issue was discussed, extensively, in numerous papers since the late 1970s (see \citealt{netzer79}). Moreover, \citet{netzer95} and \citet{bechtold97} showed that the $\mathrm{L\alpha}$/$\mathrm{H\beta}$ ratio is strongly correlated with the continuum luminosity ratio L(1215\AA)/L(4866\AA), exactly in the same manner as we find here for $\mathrm{H\alpha}$/$\mathrm{H\beta}$. These studies, and several others that followed, are all based on very small samples and will not be discussed further in this paper.

We also calculate the median $\Balmerobs$ of the NLR as a function of $\aopt$, using the narrow components of \Ha\ and \Hb\ measured by \cite{shen11}. These NLR measurements might be subjected to significant uncertainties since the NLR is weak compared to the BLR in luminous AGN and separating the two profile components is challenging and occasionally highly uncertain. The relation between $\Balmerobs$ in the NLR and $\aopt$ is shown in the right panel of Fig.~\ref{f:blamer}. The value of $\Balmerobs$ at $\aopt=\alpha_0$ is $4.07\pm0.48$, larger than the Case B value of $2.74-2.86$. This is not unexpected, since the Balmer photons are likely extincted by dust within the NLR clouds, thus increasing $\Balmerint$ above the Case B value (see fig.~6 in \citealt*{stern14}, and also \citealt{dopita02}). 
The observed slope of the relation between $\Balmerobs$ and $\aopt$ in the NLR is shallower than the expected slope derived in eqn.~(\ref{eq:dec_alpha}) and observed in the BLR. 
The weaker reddening of the NLR may be because the NLR emission line region is extended, and hence only part of the sightlines to the NLR traverse the extincting dust. This is in contrast with the BLR, which is compact and lies behind all the dust.

\subsection{Comparison with $\NNaI$ vs.\ $\ebv$ seen in local galaxies}\label{sec: other systems}

\begin{figure*}
	\center
\includegraphics[width=3.25in]{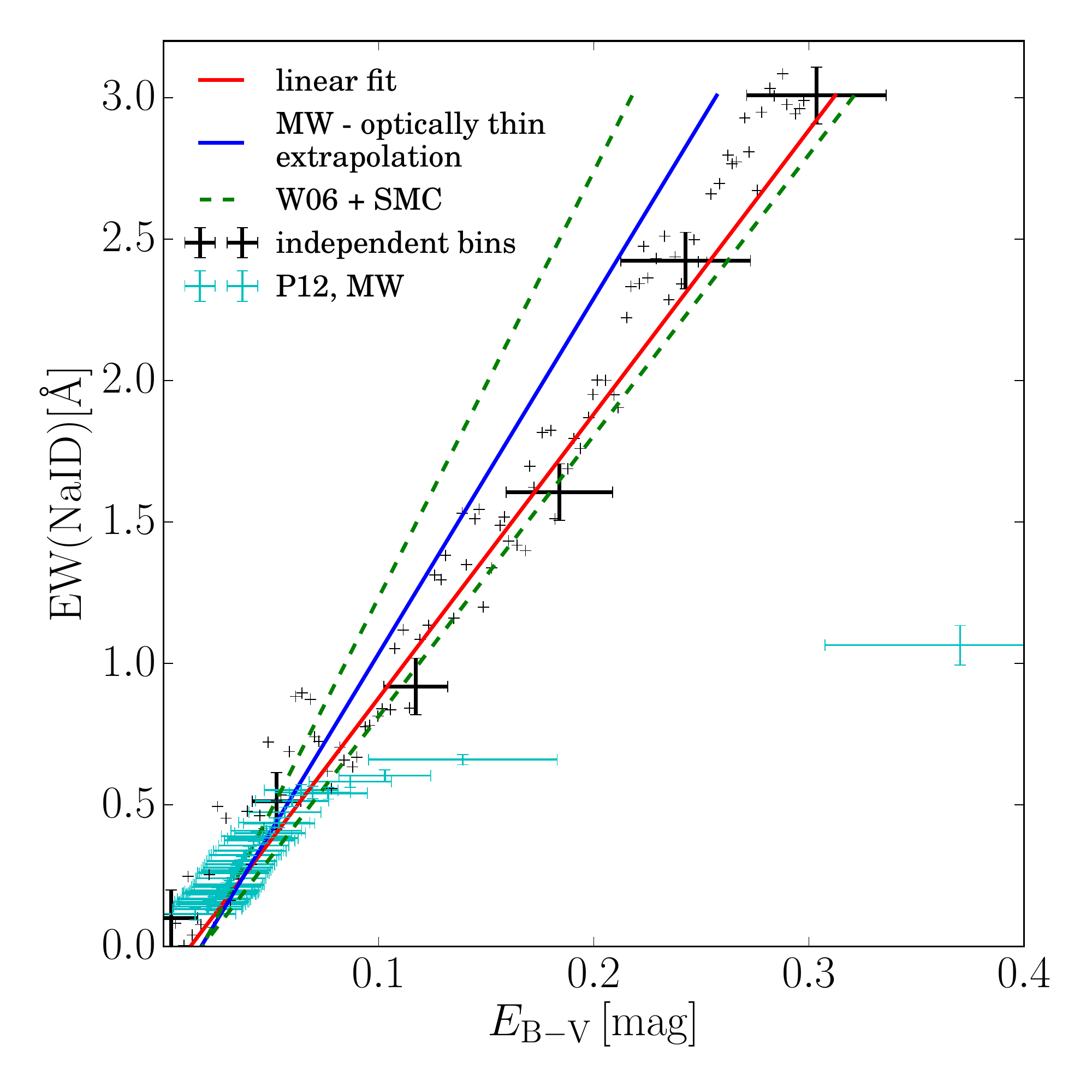}
\includegraphics[width=3.25in]{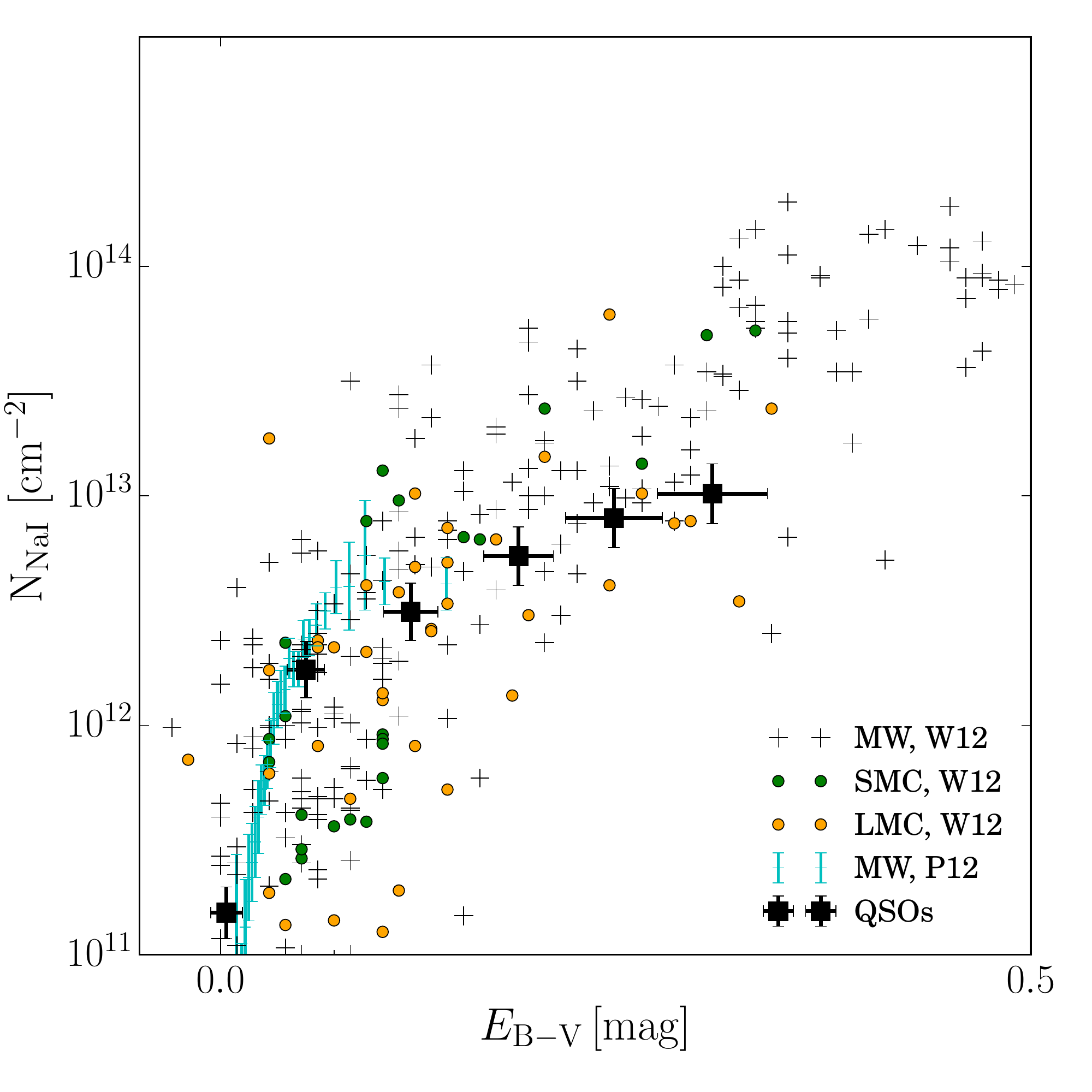}
\caption{
\textbf{(Left panel)} A comparison of the NaID EW vs.\ $\ebv$ relations in AGN, in the MW ISM, and in the Magellanic clouds ISM.
The median NaID EW for the FWHM = 2200 $\mathrm{km\,s^{-1}}$ bin in AGN measured in this worked are marked by large error bars. The running median is marked by small black crosses, while a linear fit is plotted in red. 
Cyan markers show the mean relation in the MW, in which the increase in EW with $\ebv$ saturates at EW$\sim$1\AA. 
The blue line plots the extrapolation of the MW relation in the optically thin regime to the values of $\ebv$ observed in AGN. 
The green dashed lines delinate the possible range for this relation in the SMC.
The slope of the NaID EW vs.\ $\ebv$ relation we find in AGN is within a factor of two of the relation observed in local galaxies in the optically thin regime.
\textbf{(Right panel)} A similar comparison of $\ebv$ vs.\ $\NNaI$.
The relation in AGN measured in this work is marked by black rectangles, where $\NNaI$ is calculated from EW(NaID) in the optically thin limit. 
The mean relation in the MW from P12 is plotted by cyan markers, while green circles, orange circles, and black crosses denote the individual measurements from W12 for the SMC, LMC, and MW, respectively. The relations between $\NNaI$ and $\ebv$ are similar in all shown environments. 
}
\label{f:na_ebv}
\end{figure*}

In our suggested scenario, $\aopt$ is related to $\ebv$ via eqn.~(\ref{eq:ebv_alpha}). The coefficient in eqn.~(\ref{eq:ebv_alpha}) can be calculated for an assumed extinction law, which gives 
\begin{eqnarray}\label{eq: aopt vs ebv}
 E_{\rm B-V;~MW}  &=& 0.188 \,(\alpha_0 - \aopt)  \nonumber \\
 E_{\rm B-V;~SMC} &=& 0.209 \,(\alpha_0 - \aopt)  \nonumber  \\
 E_{\rm B-V;~LMC} &=& 0.194 \,(\alpha_0 - \aopt)  \nonumber \\
 E_{\rm B-V;~GB} \,\,\,&=& 0.239 \,(\alpha_0 - \aopt)
\end{eqnarray}
where we used the prescriptions for the MW, SMC, and LMC extinction laws from \cite{pei92},  and also the 'gray dust' formulation from \citet{gaskell07}. For simplicity, we henceforth assume a MW extinction law, though the results remain the same when applying SMC, LMC, or GB extinction laws instead.
Eqn~(\ref{eq: aopt vs ebv}) implies that for $\alpha_0=-0.1$, a typical AGN with $\aopt = -0.5$ has $\ebv=0.08\mag$.

Applying eqn.~(\ref{eq: aopt vs ebv}) to the EW(NaID) vs. $\aopt$ relation found for the FWHM$=2200\kms$ bin (Table~\ref{tab: relations}) we get
\begin{equation}\label{eq: linear QSO}
 \frac{{\rm EW(NaID)_{QSO}}}{\AA} = (10.08\pm 0.55)\, \ebv
\end{equation}
This relation is seen in Figure~\ref{f:ew_alpha_all} (black dashed line).

In the left panel of Figure~\ref{f:na_ebv}, we compare equation~\ref{eq: linear QSO} to the EW(NaID) vs.\ $\ebv$ relations observed in the ISM of local galaxies. We show the measurements of Poznanski et al.\ (2012, hereafter P12), who used over a million extragalactic SDSS spectra stacked in the observer frame to study the mean relation between sodium absorption and dust extinction in the MW. 
Fig.~\ref{f:na_ebv} shows that in the MW, EW(NaID) saturates after reaching EW$\sim$1\AA, while in the AGN studied here a linear dependence between EW(NaID) and $\ebv$ persists to higher values of EW(NaID). The difference in the EW where the absorption feature saturates may be due to the difference between the integrated velocity fields along stacked MW sightlines, 
compared to the integrated velocity fields along stacked sightlines to the centers of AGN hosts.

In order to compare the P12 relation to the relation found here, we fit a linear relation to the measurements of P12 in the optically thin regime (blue line), which is:
\begin{equation}\label{eq: linear MW}
 \frac{{\rm EW(NaID)_{MW}}}{\AA} = (12.54\pm 0.35)\, \ebv
\end{equation}
One can see that this relation is close to that we find for AGN in equation \ref{eq: linear QSO}.

Additional NaID measurements for the MW and Magellanic clouds (SMC, LMC) have been compiled by \citeauthor{welty12} (2012, hereafter W12)\footnote{astro.uchicago.edu/$\sim$dwelty/coldens\_mc.html.} and contain measurements by \citet{welty06} and \citet{cox06b}.
Since the W12 measurements are given towards individual lines of sight rather than in stacked spectra, the relation exhibits a large scatter. This scatter implies an uncertainty in the $\ebv$ limit in which the optically thin regime ends. We therefore can obtain only lower and upper limits on the linear relation of EW(NaID) and $\ebv$:
\begin{eqnarray}\label{eq: linear SMC}
 \frac{{\rm EW(NaID)_{SMC}}}{\AA}_{min} = (\,\,\,9.92 \pm 0.72)\, \ebv \nonumber\\
 \frac{{\rm EW(NaID)_{SMC}}}{\AA}_{max} = (15.05\pm 0.61)\, \ebv
\end{eqnarray}
The upper and lower limit for the SMC measurements is also shown in Fig.~\ref{f:na_ebv}. This relation is consistent with the AGN relation.

In the right panel of Figure~\ref{f:na_ebv} we show the relation between $\ebv$ and $\NNaI$. We perform the comparison also in $\NNaI$ due to the difference in velocity dispersion between the different environments.  
Since there is no evidence of saturation in the AGN sample, we use the optically thin limit to calculate $\NNaI$  (e.g.~\citealt{draine11}):
\begin{equation}\label{eq: EW to NNa}
 \NNaI = 3.4\cdot 10^{12} \left(\frac{\rm EW(NaID)}{\rm \AA}\right) \cm^{-2} ~,
\end{equation}
which together with eqn.~(\ref{eq: linear QSO}) yields
\begin{equation}\label{eq: NNaI vs. ebv observed}
 \NNaI = (3.4 \pm 0.2) \cdot 10^{13}\, \frac{\ebv}{\mag} \cm^{-2} ~.
\end{equation}
For the other samples we use the $\NNaI$ published in the respective papers. 
Note that while the QSO and P12 measurements are mean relations, the Welty measurements are individual measurements which give a sense of the scatter around this mean. 
As implied by the left panel, the relation between $\NNaI$, EW(NaID) and $\ebv$ is similar in all environments shown in Fig.~\ref{f:na_ebv}.

\section{Discussion}\label{sec: discussion}

\subsection{$\NNaI$ and reddening in AGN photoionized gas}\label{sec: pi models}

Gas associated with dust along the line of sight to the central BH will be photoionized by the AGN radiation. 
We can therefore compare the observed relation between EW(NaID) and $\ebv$ to the relation expected in gas in photoionization equilibrium in the vicinity of the AGN. 

\subsubsection{The expected EW(NaID)}\label{sec: pi models ew vs ebv}

\begin{figure}
 \includegraphics{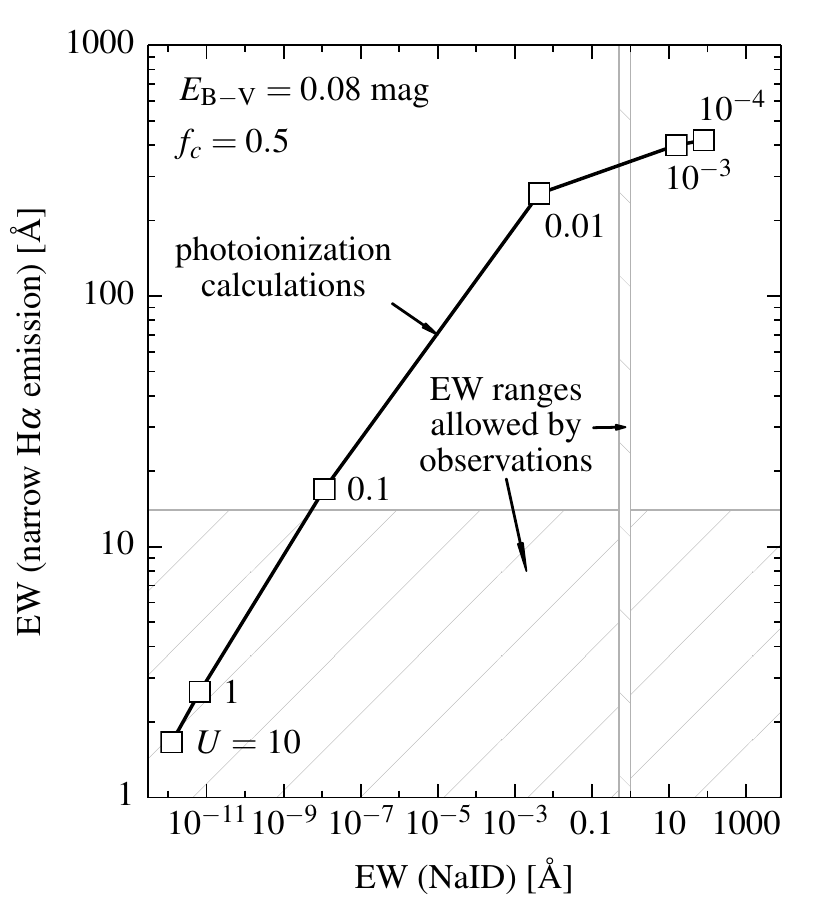}
\caption{
The expected EW(NaID) and EW(narrow \Ha\ emission) in \cloudy\ photoionization calculations of dusty-gas exposed to AGN radiation. 
The assumed dust-column of $\ebv=0.08\mag$ is appropriate for a typical AGN with $\aopt=-0.5$.
The assumed covering factor $f_c=0.5$ of the dusty gas is based on estimates that half of AGN appear as type 1s. \
The EWs are calculated for different assumptions on the ionization parameter of the dusty gas (eqn.~\ref{eq:U}).
Other model parameters are described in the text, though any plausible values for these parameters yield similar results.
The hatched regions mark the ranges of EWs allowed by observations.
The observed EW(NaID) for $\aopt=-0.5$ is based on Fig.~\ref{f:ew_alpha_all},
while the narrow \Ha\ emission from the dusty gas is assumed to be less or equal to the observed narrow \Ha\ emission. 
The model which reproduces the observed EW(NaID) ($U\approx10^{-2.5}$) over-predicts the observed \Ha\ line by a factor of $>$20, 
while models which adhere to the \Ha\ upper limit predict practically zero NaID absorption. 
This result suggests that the gas exposed to the AGN radiation does not induce the observed NaID absorption. 
}
\label{f: EW vs Ha}
\end{figure}

The ratio of the neutral sodium column $\NNaI$ to $\ebv$ is
\begin{eqnarray}\label{eq: NNaI vs. ebv theory}
 \frac{\NNaI}{\ebv} &= \fNaI \cdot \frac{\NNa}{\NH}\cdot \frac{X}{\mP} \cdot \frac{\Sgas}{\Sdust} \cdot \frac{\Sdust}{\ebv}  = 1.2\cdot10^{16}\fNaI\cdot  \nonumber \\
                   &  \left( \frac{\NNa/\NH}{2.1\cdot10^{-6}} \right)\left(\frac{\Sgas / \Sdust}{100}\right)\left( \frac{\Sdust/\ebv}{1.3\cdot10^{-4}}\right)   \cm^{-2}\, {\rm mag}^{-1} \nonumber\\
\end{eqnarray}
where $\fNaI$ is the fraction of sodium which is neutral, 
$\NNa$ is the sodium column, $\NNa/\NH$ is the sodium abundance relative to hydrogen, 
$X$ is the hydrogen mass fraction, 
$\Sgas=\NH\mP/X~[{\rm g}\,\cm^{-2}]$ is the gas surface density, 
and $\Sdust$ is the dust surface density ($\Sdust/\Sgas$ is the dust-to-gas ratio). 
The numerical values are for Galactic dust, while the sodium abundance is normalized by its solar value\footnote{\cite{weingartner01} provide circumstantial evidence that sodium is not strongly depleted in the MW ISM.}.
Comparison of eqn.~(\ref{eq: NNaI vs. ebv theory}) with eqn.~(\ref{eq: NNaI vs. ebv observed}) suggests $\fNaI \approx 0.003$. 

The value of $\fNaI$ depends mainly on the ionization parameter $U$:
\begin{equation}\label{eq:U}
 U\equiv \frac{\int_{\nu_0}^{\infty} L_\nu/ (h\nu){\rm d}\nu}{4\pi \rdust^2 n_e c} ~,
\end{equation}
where $\nu_0$ is the Lyman-edge frequency, $\rdust$ is the distance of the dusty-gas from the central radiation source and $c$ is the speed of light. 
The dependence of $\fNaI$ directly on $n_e$ or $\rdust$ (beyond the dependence on $U$) is weak. To constrain $U$ based on the observed relation, we run version 13.03 of \cloudy\ (\citealt{ferland13}) with a MW grain mixture and the following SED shape. 
The intrinsic optical-UV slope is assumed to equal the $\alpha_0=-0.1$ found above, in the range  $1100\AA<\lambda<1\mic$. The X-ray luminosity at $2\kev$ is set such that the (observed) optical to X-ray ratio is equal to typical values seen in $L=10^{45}\ergs$ AGN (\citealt{just07}). For simplicity, we interpolate with a single power-law between $1100\AA$ and $2\kev$. At $\lambda>1\mic$, $2\kev<h\nu<100\kev$, and $h\nu>100\kev$ we assume spectral slopes of $2$, $-1$, and $-2$, respectively. 
We run models with $U$ in the range $10^{-4}-10$, corresponding to $n_e = 10^4 - 0.1\, (\rdust/\kpc)^{-2}\cm^{-3}$ for a $\lbol=10^{45}\ergs$ AGN with the above SED. 
We include in the calculation the effect of radiation pressure on the dust grains, which is significant at $U\gtrsim 0.01$ (\citealt{dopita02}; \citealt*{groves04}; Stern et al.\ 2014). We use the depleted ISM abundance set implemented in \cloudy, except for the sodium abundance which we assume has solar abundance. The total gas column is set to equal $\NH = 4.4\cdot10^{20}\cm^{-2}$ which reproduces the typical $\ebv=0.08\mag$ found above. 

Figure~\ref{f: EW vs Ha} shows the EW(NaID) calculated by \cloudy\ for different $U$, compared to the observed ${\rm EW(NaID)}\approx0.8\AA$. The observed EW(NaID) is reproduced for $U\approx10^{-2.5}$. Note that only in low-ionized gas ($U \lesssim 0.01$) is any NaI absorption expected. In gas with a higher ionization level $\fNaI$ is practically zero. 
This conclusion is robust to reasonable changes in the assumed SED shape, the assumed gas metallicity, or the exact amount of sodium depletion.

\subsubsection{Line emission from the dusty gas}

Dusty gas exposed to AGN radiation is also expected to emit emission lines, which we can compare with available observations. 
Specifically, the H$\alpha$ recombination line luminosity is proportional to the recombination rate, which in turn is equal to the rate of ionizing photons absorbed by the gas. 
For a given gas column, in the optically thin limit the fraction of absorbed ionizing photons scales as $U^{-1}$, since a larger $U$ implies a smaller neutral fraction of hydrogen. In the optically thick limit ($U<10^{-2.5}$ for the assumed $\NH = 4.4\cdot10^{20}\cm^{-2}$) all the ionizing photons are absorbed by the gas and the \Ha\ luminosity does not depend on $U$. 
Some of the ionizing photons will be absorbed by the dust grains embedded in the gas. The fraction depends on $U$ and is $<50\%$ at all $U$ for the assumed $\NH$. This effect is also included in our calculation.

The \Ha\ luminosity is also proportional to the covering factor $f_c$ of the dusty gas. We estimate $f_c$ as the complement of the obscured fraction (the torus covering factor), since in our picture the dusty gas resides along all unobscured lines of sight. The obscured fraction has been estimated by comparing the volume density of type 1 and type 2 AGN (e.g.\ \citealt{reyes08}), and also by comparing the NIR-MIR luminosity of the optically thick dust with the optical-UV luminosity (\citealt{maiolino07}; \citealt*{treister08}; \citealt{mor11}; \citealt{mor12}; \citealt{roseboom13}; \citealt{lusso13}; \citealt{netzer15a}; \citealt{stalevski16}). Both types of analyses suggest $f_c \sim 0.5$ (see review in \citealt{netzer15}). 

Fig.~\ref{f: EW vs Ha} plots EW(narrow \Ha) calculated in the \cloudy\ models described above, which span the range $U=10^{-4} - 10$. Note that we approximate the distribution of $\NH$ along different lines of sight with the $\NH$ of the typical AGN. 
For comparison, we plot the range of ${\rm EW(narrow~\Ha~emission)}$ from the dusty-gas allowed by observations, which is $<14\AA$. The value of $14\AA$ is the observed EW of the narrow \Ha\ line in our sample measured by \cite{shen11}. This estimate is an upper limit on the \Ha\ emission from the line-of-sight dust, since at least some of the narrow \Ha\ emission is expected to originate from optically-thick NLR clouds.
Fig.~\ref{f: EW vs Ha} shows that only high $U>0.1$ models are consistent with the upper limit on the \Ha\ line. 
Therefore, any model that reproduces the strength of the NaID absorption implies a huge \Ha\ emission line which is inconsistent with observations, while models that are consistent with the observed narrow \Ha\ lines are inconsistent with our measured EW(NaI). 
This conclusion is independent of the exact form of the SED shape or the assumed metallicity, and is corroborated by other strong lines such as \oiii~5007\AA.

\subsection{Alternative scenario}\label{s: coherent picture}

In Fig.~\ref{f:blamer}, we showed that the dust which reddens the continuum also extincts the BLR, which suggests that the dust resides along the line of sight to the central source. 
However, Fig.~\ref{f: EW vs Ha} suggests that gas along these sightlines is highly ionized, and is not expected to create a NaI absorption feature. 
Why then are $\aopt$ and EW(NaID) correlated?

\begin{figure}
\includegraphics[width=0.5\textwidth]{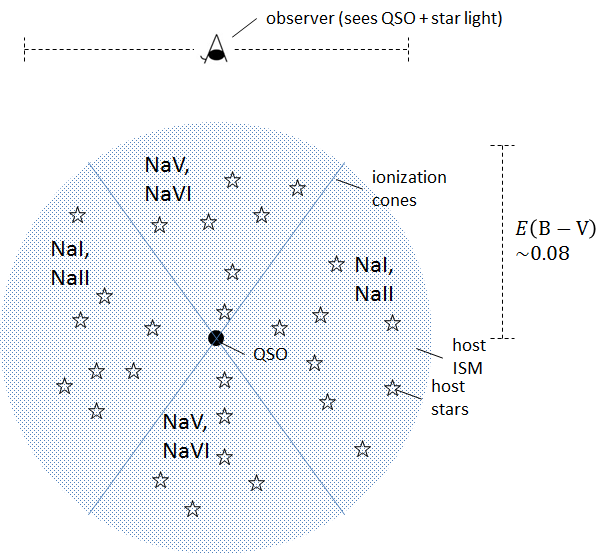}
\caption{
Our proposed schematic of a typical AGN ($\aopt=-0.5$) and its host galaxy, assumed to be spherical for simplicity. 
Small-scale obscuration shields a part (not necessarily continuous) of the host ISM from the AGN radiation.
Stellar light which passes through these regions is absorbed by NaI atoms. 
Both stellar and AGN light are extincted by the ISM dust. 
In this picture, the correlation of $\aopt$ and EW(NaID) found above is driven by the range of typical ISM columns in AGN host galaxies. 
}
\label{f: schema}
\end{figure}

One possibility is that the dusty-gas is the ISM of the host galaxy, as depicted in the schematic cartoon in Figure~\ref{f: schema}. At host galaxy scales ($\gtrsim\kpc$), some of the gas will be ionized by the AGN, but some of the gas will be shielded from the AGN radiation via obscuration on smaller scales (e.g.\ by the torus). In this picture, dust extinction occurs on all sightlines, both the sightline to the accretion disk and BLR and the sightlines to the stars. Neutral sodium however exists only in shielded regions, and hence the NaI absorption feature is imprinted only on the stellar continuum.
A correlation between $\aopt$ and EW(NaID) is then expected if ISM columns along shielded and non-shielded sightlines are correlated. Such a correlation seems plausible, for example if the range of ISM columns is driven by the inclination of our line of sight relative to the host galaxy disk. In this case, both non-shielded and shielded sightlines are expected to have larger columns in edge-on hosts compared to face-on hosts. Another possibility is that the range of ISM columns is driven by the total ISM mass. In this scenario as well, ISM-rich hosts are expected to exhibit larger columns than ISM-poor hosts along all sightlines, which would produce the observed correlation.

Is the relative stellar contribution to the continuum $\fhost$ at 6000\AA\ large enough so that the NaID absorption feature is expected to be observable?
SL12 found that in low-$z$ SDSS AGN with $\lbol=10^{45.2}\ergs$, which is the mean $\lbol$ of the objects analyzed here (Fig.~\ref{f:par_hist}), the average host emission through the SDSS 3" fiber is half of the AGN continuum emission at 6000\AA, i.e.\ $\fhost(6000\AA)=0.33$ (see fig.~13 there). 
This comparable AGN and host contributions to the continuum is further supported by the median EW(broad \Ha) of 300\AA\ in our sample (measured by \citealt{shen11} on objects in which \Ha\ is observed), compared to the mean EW(broad Ha) of $570\AA$ relative to the AGN continuum found by SL12.
For comparison, Fig.~\ref{f:stacked_example} shows that the deepest NaID absorption feature in our stacked spectra is only 10\% of the continuum emission. Therefore, it is possible that the NaID absorption feature is imprinted only on the stellar emission.

This suggested scenario naturally explains the similarity of the $\NNaI$ vs.\ $\ebv$ relation in QSOs and in local galaxies shown in Fig.~\ref{f:na_ebv}, since the conditions in the shielded ISM of AGN hosts are plausibly not very different from an inactive galaxy.
Also, this large-scale dust scenario is supported by the analysis above of the NLR Balmer ratio (Fig.~\ref{f:blamer}) and the analysis of the NLR BPT ratios vs.\ $L_{\rm UV} / L_{\rm broad~\Ha}$ in type 1 AGN (\citealt{stern13}), which both suggests that the scale of the reddening dust is at least as large as the NLR, i.e.\ on scales much larger than $100\pc$ for the luminous objects in our sample. 

The non-negligible host contribution also suggests the intrinsic AD slope is bluer than the value of $\alpha_0$ found here. 
Assuming $\fhost(3000\AA)=0.1$ and $\fhost(5100\AA)=0.25$, as suggested by \cite{shen11} and SL12 for $\lbol=10^{45.2}\ergs$ AGN\footnote{We estimate $\fhost(3000\AA)$ by interpolating the values of $\fhost(3940\AA)$ and $\fhost({\rm NUV})$ implied by the SL12 calculation. The NUV emission in the SL12 objects corresponds to an average rest frame wavelength of 1900\AA (see table 2 there).}, implies an intrinsic AD slope which is larger by $0.35$ than found above, i.e.\ an intrinsic AGN slope of $\alpha_0 +0.35=0.25$. This value is consistent with the slope of $\sim 0.2$ expected for the average $\mbh=0.7\cdot10^8\msun$ and $L/L_{\rm Edd}=0.25$ found by \cite{shen11} in objects with ${\rm FWHM}=2000\kms$. 
We note though that a more careful decomposition between the host continuum, the AGN continuum and the line emission as a function of FWHM and $\aopt$ is required to derive accurately the true intrinsic AD slope. 

\subsection{Can the range in $\fhost$ drive the EW(NaID) vs.\ $\ebv$ relation?}

In the scenario suggested in the previous section, EW(NaID) is expected to increase with $\fhost$. 
Since host galaxies have redder optical slopes than AGN, then a large $\fhost$ would also imply a relatively red $\aopt$.
One may therefore wonder whether the $\aopt$ vs.\ EW(NaID) relation is actually driven by the range in $\fhost$ in the sample, 
rather than by the distribution in the dusty gas column as suggested above. 
We find this possibility unlikely, for a few reasons. 
First, in this varying host contribution scenario one would expect the Mg\,b feature to become stronger with decreasing $\aopt$. Fig.~\ref{f:mg_stacks} suggests that there might be a weak increase in EW(Mg\,b) with $\aopt$ in the largest FWHM bins, but this trend is significantly weaker than the trend of EW(NaID) with $\aopt$, and hence the change in the host contribution with $\aopt$ implied by Mg\,b is insufficient to explain the relation of EW(NaID) with $\aopt$.
Second, a change in host contribution fails to explain why the broad \Ha/\Hb\ ratio changes with $\aopt$ (Fig.~\ref{f:blamer}). And third, the required change in $\fhost$ should be evident as a change in EW(broad \Ha) with $\aopt$, in contrast with the reddening scenario where EW(broad \Ha) is expected to remain constant. 
In our sample, objects with $\aopt=-0.4$ have a median EW(broad \Ha)$=340\AA$ which suggests $\fhost(6560\AA) \approx 1-340/570 = 0.4$,
while objects with $\aopt=-1.5$ have a median EW(broad \Ha)$=270\AA$ which suggests $\fhost(6560\AA) \approx 1-270/570 = 0.52$.
Therefore, $\fhost$ likely increases by a mild factor of $\sim1.3$ from objects with $\aopt=-0.4$ to objects with $\aopt=-1.5$. 
This increase in $\fhost$ falls short of the increase in the median EW(NaID) by a factor of $3.6$, from $1.1\AA$ at $\aopt=-0.4$ to $4.0\AA$ at $\aopt=-1.5$\footnote{The values of $\aopt=-0.4$ and $\aopt=-1.5$ are chosen for this test to avoid the difference in FWHM distribution implied by using objects with $\aopt>-0.4$ or $\aopt<-1.5$ (see Fig.~\ref{f:ew_alpha_all}).}. Therefore, the change in $\fhost$ implied by the change in EW(broad H$\alpha$) is too weak to explain the EW(NaID) vs.\ $\aopt$ relation.

We note that a correlation between $\aopt$ and $\fhost$ is also expected if the physical size spanned by the SDSS fiber changes with $\aopt$. However, no such trend exists in our sample. Objects with $\aopt=-2$ have a mean $z$ of $0.41$, similar to the mean $z=0.435$ of objects with $\aopt=0$.

\subsection{EW(NaID) and $\alpha_0$ vs.\ FWHM(H$\beta$)}\label{s: EW offset}
\newcommand{\rblr}{R_{\rm BLR}}
What is the origin of the offset between the EW(NaID) vs.\ $\ebv$ relations in different FWHM bins (Fig.~\ref{f:ew_alpha_all})?
Assuming the BLR is in virial motion, then $\mbh \propto {\rm FWHM}^2 \rblr$, where $\rblr$ is the size of the BLR. Reverberation mapping studies have shown that $\rblr \propto L^{0.6 \pm 0.1}$ (\citealt{kaspi05, bentz09}). Therefore, in our sample where the dynamical range in $L$ is small (Fig.~\ref{f:par_hist}), we expect $\mbh$ to increase with FWHM. The different implied $\mbh$ in the different FWHM-bins may suggest a different intrinsic AGN SED, which can change $\alpha_0$. 

Another possible difference between the different FWHM bins arises from the the well-known relations between $\mbh$ and the bulge properties (\citealt{magorrian98, ferrarese00}), which have been shown to apply also to the host galaxies of AGN (\citealt{laor98, peterson04}). The typical $\fhost$ and/or stellar age may therefore change with FWHM. This conclusion is supported by Fig.~\ref{f:mg_stacks}, which shows that absorption features which arise in stellar photospheres become more prominent at large FWHM (see also fig.~18 in SL12). 

\subsection{Implied $\ebv$ distribution}

\begin{figure}
 \includegraphics{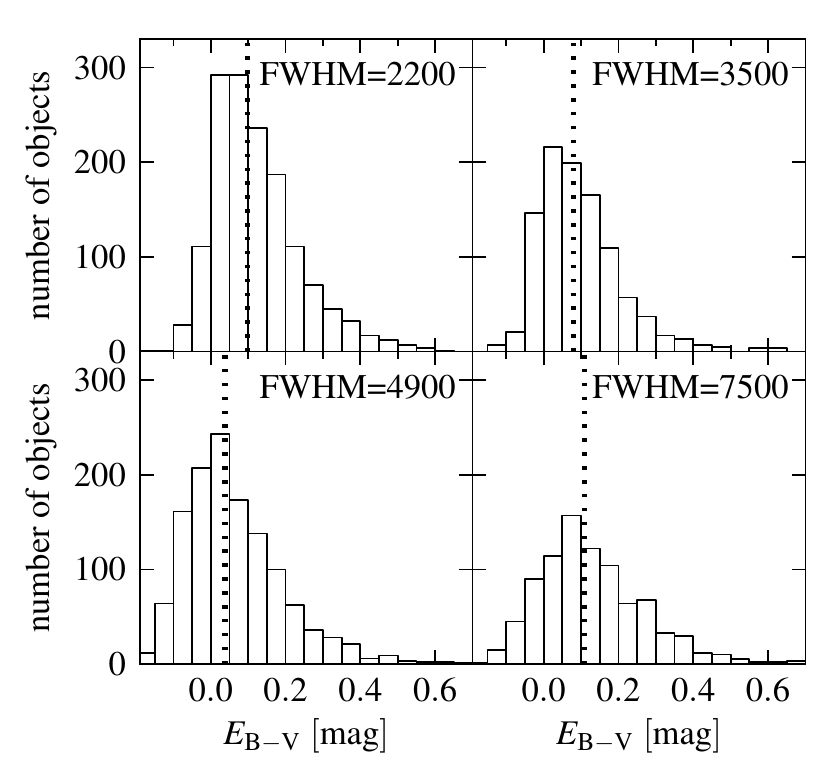}
\caption{
The implied $\ebv$ distributions for different FWHM bins (noted in km s$^{-1}$) in the AGN sample. The value of $\ebv$ is derived from $\aopt$ using $\ebv = 0.2\, (\alpha_0 - \aopt)$ (eqn.~\ref{eq: aopt vs ebv}), and assuming the $\alpha_0$ listed in Table~\ref{tab: relations}. The median $\ebv$ in each panel is marked by a dotted line. The entire sample has a mean $\ebv=0.1\mag$ and a median $\ebv=0.08\mag$. 
}
\label{fig: ebv distribution}
\end{figure}

The measured $\aopt$ in our sample can be converted to $\ebv$ using eqn.~(\ref{eq: aopt vs ebv}). 
The top-left panel in Figure~\ref{fig: ebv distribution} shows the $\ebv$ distribution of AGN with FWHM=$2200\kms$, assuming a MW extinction law (values are larger by 10\% for an SMC law). The mean $\ebv$ is 0.12 mag, while the median $\ebv$ is 0.1 mag.
Note that some QSOs have negative $\ebv$ due to slopes bluer than the derived $\alpha_{0}$. The standard absolute deviation of these negative $\ebv$ values is $0.039$ mag.  These negative values provide an estimate of the various systematic errors in our derivation of $\ebv$, probably due to variance in the host and BLR contribution to $\aopt$, and due to the dispersion in the intrinsic AD slope. If this scatter is solely due to the distribution in intrinsic slopes, it suggests a standard deviation of 0.2 in the intrinsic slope distribution.

The similarity of the slopes of the EW(NaID) vs. $\aopt$ relations in all FWHM bins (see table \ref{tab: relations} and Fig.~\ref{f:ew_alpha_all}), suggests that we can use the same relation to derive $\ebv$ also in the bins with FWHM$=3500$, $4900$ and $7500\kms$. However, the determination of $\alpha_0$ in these large FWHM bins is not straightforward, since in these bins there are no stacks with ${\rm EW(NaID)}=0$, possibly due to the contribution to the NaID absorption feature from stellar photospheres. 

Assuming that $\alpha_0$ is equal to the lowest $\aopt$ seen for each FWHM (listed in Table~\ref{tab: relations}), implies a mean $\ebv=0.1\mag$ for the entire sample, and a median $\ebv=0.08\mag$. 
The implied $\ebv$ distributions under this assumption are shown in Fig.~\ref{fig: ebv distribution}. 
Alternatively, assuming that $\alpha_0$ at all FWHM bins is equal to the $\alpha_0=-0.1$ found for FWHM$=2200\kms$, implies a somewhat larger mean $\ebv=0.13\mag$ and median $\ebv=0.1\mag$. 
In the extreme case where $\alpha_0$ is equal to the extrapolation of the EW(NaID) vs. $\aopt$ relation to ${\rm EW(NaID)}=0$ ($\alpha_0=0.18,\ 0.37,\ 0.71$ for FWHM$=3500$, $4900$ and $7500\kms$, respectively), the implied mean $\ebv$ is $0.19\mag$ and the median $\ebv$ is $0.17\mag$. 

To summarize, at least half of the AGN in our sample have $\ebv > 0.08\mag$, and at least a quarter have $\ebv > 0.18\mag$.
For comparison, 
\citet{lusso13} found that only 24\%\ of type~1s in the XMM-COSMOS survey have $\ebv > 0.1\mag$. 
\citet{krawczyk15} found that 2.5\% (13\%) of the non-BAL (BAL) in their SDSS-based sample have $\ebv > 0.1\mag$, and 0.1\% (1.3\%) have $\ebv > 0.2\mag$. 
\citet{capellupo15} found that 70\% of their sample have negligible dust columns.
The lower dust columns found by previous studies compared to this work (except for \citealt{capellupo15}) are a direct result of their assumption that the typical AGN is not reddened by dust.

Using the cloudy models described above, we obtain the fraction of incident radiation absorbed by the dusty gas for the typical $\ebv=0.08\mag$ AGN ($\aopt=-0.5$). We find that given the assumed SED and grain properties, about a third of the incident, radiation, integrated over all wavelengths, is absorbed by the dust along a typical line of sight. This result suggests that the bolometric luminosity $\lbol$ of type 1 AGN, as derived from the optical, are typically underestimated by a factor of $(2/3)^{-1}=1.5$.

\subsection{Expected FIR and X-ray properties}

Given the very high level of ionization of the newly suggested line-of-sight gas,  the corresponding column of dust, and the large covering factor, we need to consider other types of emission and absorption by this component.

The \cloudy\, model described above can be used to estimate the fraction of the radiation absorbed by the dusty line-of-sight component for every value of $\ebv$ and covering factor. As before, we do not consider a component which is less than about $100\pc$ from the center since our model assumes similar column densities on galactic scales. 
As already noted, for the typical $\ebv = 0.08\mag$ case, approximately a third of the incident radiation is absorbed
by the dust. Given the assumed distance, this energy must be re-radiated at far infrared (FIR) energies with peak wavelength that depend on the dust location. For ISM-type densities ($\sim 1$ cm$^{-3}$) and a typical bolometric luminosity of $10^{45}\ergs$, the required $U$ of $\sim1-10$ (Fig.~\ref{f: EW vs Ha}) is achieved at $1-10\kpc$. The corresponding dust temperatures, assuming optically thin dust and gray body emission with spectral emissivity index $\beta = 1.5-2$, is in the range $10-40\K$. Given the fraction absorbed and a typical covering factor of $\sim 0.5$, the total luminosity emitted in the FIR is hence of order $0.15\,\lbol$. The FIR emission in individual objects may vary significantly around this average value, depending on the $\ebv$ distribution along different directions in single sources which could be different than the $\ebv$ distribution seen in a sample. However, the ensemble average FIR emission is very large and such emission cannot escape detection. We note that there are also other claims in the literature for cold dust in AGN (e.g., \citealt{symeonidis16}), but these results are still open to interpretation.

It is not our intention to investigate the issue of FIR emission by the obscuring dusty gas in this paper. Such a study requires more detailed considerations, comparison with other FIR sources of emission (mostly star formation in the host galaxy), and more. We will return to this topic in a forthcoming publication.

The consideration detailed above, that are summarized in Fig.~\ref{f: EW vs Ha}, suggest $U\sim 1-10$. Such a large ionization parameter is typical of X-ray gas in AGN \citep{netzer13}, and may have observational consequences that have not been studied so far. For example, if $U\sim 10$ and the UV-X-ray SED is typical of AGN with $\lbol=10^{45}\ergs$, elements like O, Ne, Mg and Si will be very highly ionized. In this case, the strongest X-ray features are absorption and emission lines of OVIII, NeX, MgXII, etc. Emission line intensities, absorption line EWs, and other properties depend, again, on the gas distribution, velocity field, and level of ionization at different locations. This topic is beyond the scope of the present paper and is
differed to our future paper. 

\section{Conclusions}\label{s:conc}

We use 4\,946 $\lbol\approx10^{45}\ergs$ QSOs at $z\sim0.4$ from the SDSS and bin them by their optical slope $\aopt$ and FWHM of the broad $\mathrm{H}\beta$ line. We measure $\aopt$ and the EW of the NaID absorption doublet on the stacked spectrum of each bin. Our results can be summarized as follows:

\begin{enumerate}
	\item For each bin in FWHM, we find a linear relation between $\aopt$ and EW(NaID), such that EW(NaID) increases as $\aopt$ becomes smaller (redder continuum). The slopes of these relations are similar in all FWHM bins. In the smallest $1000<{\rm FWHM}<3000\kms$ bin we find ${\rm EW(NaID)}=0$ at $\aopt=-0.1$. A similar trend is not seen in the Mgb photospheric absorption feature, suggesting that the trend in NaI has an interstellar origin.

       \item The blue slope corresponding to ${\rm EW(NaID)}=0$ is similar to that predicted for a thin accretion disk after allowing for the host galaxy contribution. This result is consistent with a scenario where the observed $\aopt$-distribution in AGN is driven by the column of dust along the line of sight, as previously suggested by several studies. 
       The relationship between slope and reddening by dust is given by $\ebv \approx 0.2\cdot(-0.1 - \aopt)$. The reddening of a typical type 1 AGN with $\lbol=10^{45}\,\ergs$ and $\aopt=-0.5$ is $\ebv\approx0.08$ mag, suggesting that AGN bolometric luminosities are typically underestimated by $\approx 40\%$. 

       \item The ubiquitous reddening dust interpretation is further supported by the observed relation between $\aopt$ and the BLR $\Ha/\Hb$ ratio. In unreddened objects the mean line ratio is $\Ha/\Hb \sim 3$ similar to the result of \cite{dong08}, though with a broad distribution in the range $1.5-4$ which suggests that simple Case B recombination cannot explain the ratio in all sources.

		\item Photoionization calculations show that neutral sodium does not exist in appreciable quantities in gas exposed to the AGN radiation. We therefore suggest that the dusty gas is the ISM of the host galaxy, and that the NaID absorption feature arises in regions shielded from the AGN radiation, along the line of sight to the stars. The correlation of $\aopt$ with EW(NaID) hence arises from the plausible assumption that the ISM columns along shielded and non-shielded sightlines are correlated. This suggested scenario is supported by the similarity of the $\ebv$ vs.\ NaI column relation found here and the same relation in the Milky Way found by \citet{poznanski12}. Therefore, our results suggest that the observed optical-UV continuum slopes of quasars provide a novel constraint on the ISM of their host galaxies, which could be useful for example for studies simulating AGN feedback. Other observational consequences, that are not discussed in the present work, are significant FIR emission and X-ray absorption by the new component.
	   
\end{enumerate}
	
\acknowledgments 
\vspace{.5cm}
We thank the anonymous referee whose remarks and suggestions helped improve this manuscript.
JS acknowledges financial support from the Alexander von Humboldt foundation. 
DP is partially supported by Grant No 2014413 from the United States-Israel Binational Science Foundation (BSF). 
HN acknowledges support by the Israel Science Foundation (ISF) grant 284/13.
We thank 
N. Arav,
J.~F. Hennawi,
A. Laor,
B. Trakhtenbrot, 
J.~X. Prochaska, and 
D. Welty for useful discussions regarding this work and manuscript. 

The bulk of our computations was performed on the resources of the National Energy Research Scientific Computing Center, which is supported by the Office of Science of the U.S. Department of Energy under Contract No. DE-AC02-05CH11231. The  spectroscopic analysis was made using IPython \citep{perez07}. We also used the following Python packages:  pyspeckit\footnote{www.pyspeckit.bitbucket.org}, healpy\footnote{www.healpy.readthedocs.org}, and astropy\footnote{www.astropy.org/}.

This work made extensive use of SDSS-III\footnote{www.sdss3.org} data. Funding for SDSS-III has been provided by the Alfred P. Sloan Foundation, the Participating Institutions, the National Science Foundation, and the U.S. Department of Energy Office of Science. SDSS-III is managed by the Astrophysical Research Consortium for the Participating Institutions of the SDSS-III Collaboration including the University of Arizona, the Brazilian Participation Group, Brookhaven National Laboratory, Carnegie Mellon University, University of Florida, the French Participation Group, the German Participation Group, Harvard University, the Instituto de Astrofisica de Canarias, the Michigan State/Notre Dame/JINA Participation Group, Johns Hopkins University, Lawrence Berkeley National Laboratory, Max Planck Institute for Astrophysics, Max Planck Institute for Extraterrestrial Physics, New Mexico State University, New York University, Ohio State University, Pennsylvania State University, University of Portsmouth, Princeton 
University, the Spanish Participation Group, University of Tokyo, University of Utah, Vanderbilt University, University of Virginia, University of Washington, and Yale University.

\bibliographystyle{aasjournal}
\bibliography{ref}

\end{document}